\documentclass[]{AIAA}		% for proceedings (20-page limit)

 \usepackage{lettrine}% dropped capital at beginning of paragraph
\usepackage{bm}
\usepackage{amsmath}
\usepackage{amssymb}
\usepackage{subfigure}
\usepackage{array}
\usepackage{multirow}
\usepackage[]{natbib}
\usepackage{bm}
\usepackage{soul}
\usepackage{algorithm}
\usepackage{algorithmic}
\usepackage{xcolor}
\usepackage{multirow}
\usepackage{placeins}

\DeclareMathOperator*{\argmin}{arg\,min}

\begin{document}

\title{Stochastic Expansions Including Random Inputs on the Unit Circle}

\author{Brandon A. Jones\footnote{Assistant Professor, Department of Aerospace Engineering and Engineering Sciences, C0600, AIAA Senior Member.}}
\affiliation{University of Texas at Austin, Austin, TX, 78712, USA}
\author{Marc Balducci\footnote{Graduate Research Assistant, Colorado Center for Astrodynamics Research, UCB 431.}}
\affiliation{University of Colorado, Boulder, CO, 80309, USA}

\begin{abstract}
Stochastic expansion-based methods of uncertainty quantification, such as polynomial chaos and separated representations, require basis functions orthogonal with respect to the density of random inputs.  Many modern engineering problems employ stochastic circular quantities, which are defined on the unit circle in the complex plane and characterized by probability density functions on this periodic domain.  Hence, stochastic expansions with circular data require corresponding orthogonal polynomials on the unit circle to allow for their use in uncertainty quantification.  Rogers-Szeg\H{o} polynomials enable uncertainty quantification for random inputs described by the wrapped normal density.  For the general case, this paper presents a framework for numerically generating orthogonal polynomials as a function of the distribution's characteristic function and demonstrates their use with the von Mises density.  The resulting stochastic expansions allow for estimating statistics describing the posterior density using the expansion coefficients.  Results demonstrate the exponential convergence of these stochastic expansions and apply the proposed methods to propagating orbit-state uncertainty with equinoctial elements.  The astrodynamics application of the theory improves robustness and accuracy when compared to approximating angular quantities as variables on the real line.
\end{abstract}

\maketitle

\section*{Nomenclature}
\noindent\begin{tabular}{@{}lcl@{}}
$(a,h_e,k_e,p_e,q_e,\lambda)$ &=& equinoctial orbital elements \\
$\mathcal{A}$ &=& stochastic ordinary differential operator\\
$a$ &=& semimajor axis (one of the equinoctial elements) \\
$\mathbb{C}$ &=& space of complex numbers \\
$\bm{C}$ &=& matrix comprised of expansion coefficients\\
$\bm{c}_{\bm{\alpha}}$ &=& vector of polynomial chaos expansion coefficients\\
$d$ &=& number of random inputs\\
$\mathbb{E}[x]$ &=& expected value of $x$\\
$\mathcal{I}(x)$ &=& imaginary part of complex number $x$ \\
$I_n$ &=& modified Bessel function of order $n$ \\
$i$ &=& imaginary number $\sqrt{-1}$ \\
$J$ &=& cost function \\
$J_2$ &=& measure of equatorial gravity perturbation \\
$\mathcal{N}(\mu,\sigma^2)$ &=& Gaussian/normal distribution with mean $\mu$ and variance $\sigma^2$\\
$\mathbb{N}$ &=& space of natural numbers\\
$M$ &=& number of surrogate training samples \\
$\mathcal{P}$ &=& probability measure \\
$P$ &=& cardinality of $\Lambda_{p,d}$\\
$p$ &=& maximum degree of terms in polynomial expansion\\
$q$ &=& parameter for Rogers-Szeg\H{o} polynomials \\
$\mathbb{R}$ &=& space of real numbers \\
$\mathcal{R}(x)$ &=& real part of complex number $x$ \\
$r$ &=& rank of separated representation \\
$s$ &=& weighting coefficients in separated representation \\
$\mathbb{T}$ &=& space of complex numbers on unit circle
\end{tabular}

\noindent\begin{tabular}{@{}lcl@{}}
$T_n$ &=& Toeplitz matrix\\
$t$ &=& time \\
$\bm{u}$ &=& vector of quantities of interest \\
$\widehat{u}$ &=& estimated value of variable $u$\\
$U$ &=& matrix of propagated training samples \\
$\mathcal{VM}(\mu,\kappa)$ &=& von Mises distribution with location $\mu$ and concentration $\kappa$\\
$\mathcal{WN}(\mu,\sigma^2)$ &=& wrapped normal distribution with mean $\mu$ and variance $\sigma^2$\\
$\bm{\alpha}$ &=& multi-index in $\mathbb{N}^d_0$\\
$\Gamma^d$ &=& $d$-dimensional hypercube - the image of random variables $\bm{\xi}$\\
$\delta_{ij}$ &=& (discrete) Dirac delta function\\
$\epsilon$ &=& relative error/difference \\
$\eta_n$ &=& Verblunsky coefficient \\
$\kappa$ &=& von Mises distribution concentration parameter\\
$\lambda$ &=& mean longitude (one of the equinoctial elements) \\
$\Lambda_{p,d}$ &=& subset of multi-indices specifying tensor product of maximum degree $p$ and dimension $d$\\
$\mu_x$ &=& mean of random variable $x$ \\
$\mu_\oplus$ &=& gravitation parameter of the Earth\\
$\bm{\xi}$ &=& vector of independent random variables $\xi_i$\\
$\rho(\bm\xi)$ &=& joint probability measure\\
$\Sigma$ &=& Borel sigma algebra for probability space\\
$\sigma_x$ &=& standard deviation of random variable $x$\\
$\phi_n$ &=& characteristic function of order $n$ \\
$\Psi$ &=& matrix of evaluated multi-variate orthogonal polynomials \\
$\psi_{\alpha_i}$ &=& univariate polynomial of degree $\alpha_i$ \\
$\psi_{\bm{\alpha}}$ &=& multi-variate polynomial defined by multi-index $\bm{\alpha}$\\
$\Omega$ &=& event space
\end{tabular}

\maketitle{} 		

%=============================================================================================
\section{Introduction}

Modern engineering problems require methods of tractable uncertainty quantification for problems with data, both inputs and outputs, in a variety of domains.  For example, position is represented as a quantity on the real line.  Attitude states may be parameterized via one of several methods (quaternions, Euler angles, etc.) that are defined over a domain that is periodic over one (or more) independent variables.  The simplest form of these are circular data defined on the unit circle, and, hence, are periodic over $[0,2\pi)$ (or an equivalent domain).  Using methods of stochastic expansions for uncertainty quantification for problems that include circular data (or the higher-dimension analogues on the sphere, etc.) require an extension to existing theory to account for their differences when compared to data on the real line.  This paper describes an approach to uncertainty quantification via stochastic expansions, namely Polynomial Chaos Expansions (PCEs) and Separated Representations (SR), when including inputs and outputs defined on the unit circle.

PCEs and SR provide a means for approximating the solution of a stochastic ordinary differential equation that is square-measurable, possibly non-Gaussian, with respect to the input uncertainties. Such techniques use a projection of the stochastic solution onto a basis of orthogonal polynomials in stochastic variables that is dense in the space of finite-variance random variables. To maintain efficiency in the expansion, the basis functions must be orthogonal with respect to the density function of the inputs.  In cases of random variables that fit in the Wiener-Askey scheme, e.g., Hermite polynomials for Gaussian random inputs, selection of the basis is straightforward.  However, previous research fails to identify the correct polynomial basis for random inputs on the unit circle.  

A principle motivation of this work is in the context of astrodynamics problems.  Orbit state uncertainty propagation is an active area of research, and the potential issues were first raised in~\cite{junkins_1996}.  Proposed methods for representing and propagating uncertainty include state transition tensors~\citep{fujimoto_2012b}, Gaussian mixtures~\citep{demars_2013a,horwood_2011}, polynomial chaos~\citep{jones_2013a}, separated representations~\citep{balducci_2017}, differential algebra~\citep{valli_2013}, and others.  Applications of stochastic expansion-based methods to astrodynamics have been demonstrated for both Earth-centered and interplanetary missions \citep{jones_2013a,jones_2013b,jones_2015b,balducci_2017,feldhacker_2016d}.  They also enable robust and reliable design via optimization under uncertainty \citep{feldhacker_2016d}.  To date, most demonstrations of stochastic expansions to astrodynamics problems focus on Cartesian coordinates, and some work is required to rigorously apply them to uncertainty propagation with orbit element sets.

In \cite{junkins_1996}, the authors demonstrated the use of classical orbital elements for improved uncertainty propagation, and the equinoctial elements have grown in popularity since they reduce complications with singularities.  The various forms of the equinoctial elements include an angle describing location of the spacecraft in its orbit (e.g., see \citep{hintz_2008}).  To account for circular data in the formulation of the state Probability Density Function (PDF) with the equinoctial elements, \cite{horwood_2014} proposes a Gauss-von Mises density.  To enable an equinoctial elements-based approach with stochastic expansions and, more generally, any problem with circular data, this paper describes how to formulate basis functions on the unit circle for the polynomial methods.  Simulated results demonstrate the rapid convergence when using the appropriate orthogonal basis, and leverages the approach for orbit-state uncertainty propagation with equinoctial elements.

%The equinoctial element set includes one angle variable, which has a probability density defined on the unit circle.  The application of stochastic expansion-based methods remains largely unexplored for such cases.  To maintain efficiency in the expansion, the basis functions must be orthogonal with respect to the density function of the inputs.  In cases of random variables that fit in the Wiener-Askey scheme, e.g., Hermite polynomials for Gaussian random inputs, selection of the basis is straightforward.  However, previous research fails to identify the correct polynomial basis for random inputs on the unit circle.  In the context of orbit state uncertainty propagation, this prevents the use of the appropriate basis for random inputs associated with the angle quantity in the equinoctial elements.

This paper presents a framework for including circular data and random inputs in stochastic expansions, and demonstrates their use for orbit-state uncertainty propagation.  For circular data with probability density described by the Wrapped Normal Distribution (WND), the Roger-Szeg\H{o} polynomial provide an orthogonal basis for use in stochastic expansions.  Use of these polynomials requires recasting the random inputs into complex variables, and the paper discusses the implications of this in the context of surrogate approximations.  For random inputs not described by the WND, a general framework is presented and numeric results use the proposed procedure with the circular random inputs characterized by the von Mises density.  The paper includes a demonstration of rapid convergence (as a function of polynomial degree) of the mean and variance of the posterior PDF for a simple system.  Results when propagating orbit-state uncertainty via equinoctial elements is then presented as a more practical application of the theory.

The paper outline is as follows.  Section~\ref{sec:stoch_expansions} states the problem formulation and describes the (general) use of stochastic expansions as a solution to the problem.  Section~\ref{sec:random_inputs_uc} describes statistical descriptions of circular data and the orthogonal polynomials on the unit circle for use in stochastic expansions.  Section~\ref{sec:stoch_expansion_solution} describes their use in propagation of uncertainty via stochastic expansions, and Section~\ref{sec:numerical_examples} then demonstrates the performance of the approach.  Finally, conclusions are discussed.

%=============================================================================================
\section{Stochastic Expansions}\label{sec:stochastic_expansions}\label{sec:stoch_expansions}

\subsection{Problem Setup}\label{sec:problemsetup}

Let $(\Omega,\Sigma,\mathcal{P})$ be a probability space with event space $\Omega$, sigma algebra $\Sigma$, and probability measure $\mathcal{P}$.  Let $\bm{\xi}\in\mathbb{R}^d:\Sigma\rightarrow\Gamma^d\subset\mathbb{C}^d$ defined on the probability space be random inputs with stochastic dimension $d$.  Assume that the elements of $\bm{\xi}$ are independent but not necessarily identically distributed.  Methods based on stochastic expansions seek to produce an approximate solution to the stochastic Ordinary Differential Equation (ODE) problem
\begin{align}
	\mathcal{A}(t,\bm{\xi};\bm{u}) = 0,\hspace{18pt}(t,\bm{\xi})\in[0,t_f]\times\Gamma^d,\label{eqn:stochastic_ode}
\end{align}
where $\mathcal{A}$ is the ODE operator, $\bm{u}$ are the quantities of interest (e.g., propagated position and velocity), and $t\in[0,t_f]$ is time.  Random inputs $\bm{\xi}$ may correspond to stochastic initial conditions and/or force model parameters.  

This work considers the case where at least one element of $\bm{\xi}$ is defined on a circle.  Such quantities may be parameterize as angles, unit vectors, or by the circular variable $z\in\mathbb{T}$ where
\begin{align}
	\mathbb{T} \equiv \{ z\in\mathbb{C}\,:\,|z|=1\},
\end{align}
i.e., numbers on the unit circle in the complex plane.  Euler's formula
\begin{align}
	z = e^{i\lambda},
\end{align}
where $i=\sqrt{-1}$, relates the circular variable $z$ with the angle $\lambda$.  Such circular data have a PDF $\rho(\lambda)$ defined on the unit circle and account for periodicity on the domain, e.g., $\rho(\lambda) = \rho(\lambda+2\pi)$.  To leverage theory describing directional quantities on $\mathbb{T}$, the codomain of the expansions described in this paper are defined in the space of complex numbers.  The following sections present a method for including random inputs corresponding to directional quantities in stochastic expansions.  Note that the presentation of stochastic expansion methods is generalized to admit elements of $\bm{u}$ defined as complex numbers to accommodate such Quantities Of Interest (QOIs).  Note that we use $\lambda$ as a generic angle variable for the sake of consistency between presented theory and future application to orbit-state propagation.

%-----------------------------------------------------------------------------------------------------
\subsection{Polynomial Chaos Expansions}

A Polynomial Chaos Expansion (PCE) provides a means for generating an approximate solution to the stochastic ODE problem in (\ref{eqn:stochastic_ode}) by projecting the solution onto a dense, multi-variate polynomial basis on the space of random inputs $\bm{\xi}$ and orthogonal with regards to the density $\rho(\bm{\xi})$.  This method was first proposed by \cite{wiener_1938} for Gaussian random inputs, extended on by \cite{ghanem_2002,ghanem_1998,ghanem_1999a,ghanem_1999}, and later applied to a larger class of random inputs based on the Wiener-Askey scheme \citep{xiu_2002a}.  Recent work focuses on extending such methods to a broader scope of problems with high dimension \citep{doostan_2007b,doostan_2009,ma_2009,nouy_2010,doostan_2011,yang_2012,doostan_2013}, computationally expensive simulation software (see discussion and motivation in \cite{narayan_2012,ng_2014,zhu_2014,narayan_2014}), nonlinear dynamics over long time spans \citep{gerritsma_2010,wan_2005,wan_2006,jones_2013a,jones_2013b,jones_2015b}, and/or discontinuities in the solution as a function of the random inputs $\bm{\xi}$ \citep{wan_2005,wan_2006,peng_2015}.

A PCE approximates the solution for a quantity of interest vector $\bm{u}(t,\bm{\xi})$ via
\begin{align}
	\bm{u}(t,\bm{\xi}) \approx \widehat{\bm{u}}(t,\bm{\xi}) = \sum_{\bm{\alpha}\in\Lambda_{p,d}} \bm{c}_{\bm{\alpha}}(t)\,\psi_{\bm{\alpha}}(\bm{\xi})\label{eqn:general_pce_equation}
\end{align}
where
\begin{align}
	\bm{c}_{\bm{\alpha}}(t) &= \int \bm{u}(t,\bm{\xi}) \psi_{\bm{\alpha}}(\bm{\xi}) \rho(\bm{\xi})\,d\bm{\xi},\\
	\Lambda_{p,d} &\equiv \left\{\bm{\alpha} \in\mathbb{N}^d_0\,:\,\left\Vert\bm{\alpha}\right\Vert_1\leq p,\left\Vert\bm{\alpha}\right\Vert_0\leq d\right\},
\end{align}
$p$ is the maximum degree of the polynomial expansion, and the multi-variate basis functions $\psi_{\bm{\alpha}}$ are defined by the density $\rho(\bm{\xi})$ (see Section~\ref{sec:basis_functions}).  For the sake of compact notation, dependence on time is hereafter removed.  Elements of the set $\Lambda_{p,d}$ are multi-indices that define the multi-variate polynomial via
\begin{align}
	\psi_{\bm{\alpha}}(\bm{\xi}) = \prod_{j=1}^d\psi_{\alpha_j}(\xi_j),
\end{align}
i.e., the multi-variate basis function is a tensor product of univariate basis functions each of degree $\alpha_j\in\bm{\alpha}$ and a function of a single random variable $\xi_j$, and 
\begin{align}
	P \equiv |\Lambda_{p,d}| = \frac{(p+d)!}{p!d!}
\end{align}
is the number of terms in the PCE.

This work primarily leverages sampling-based (non-intrusive) methods for approximating the coefficients $\bm{c}_{\bm{\alpha}}$, which treats an existing, deterministic ODE solver as a black box.  Let $\{\bm{\xi}_m\}_{m=1}^M$ be the set of $M$ random input vectors, and $\bm{u}(\bm{\xi}_m)$ the quantity of interest generated using $M$ evaluations of the prescribed ODE solver.  The least-squares approach to solving for a PCE minimizes the least-squares cost function (e.g., see \cite{hosder_2006})
\begin{align}
	J(\bm{c}_{\bm{\alpha}}) = \sum_{m=1}^M \left(\widehat{\bm{u}}(\bm{\xi}_m;\bm{c}_{\bm{\alpha}}) - {\bm{u}}(\bm{\xi}_m;\bm{c}_{\bm{\alpha}})\right)^H \left(\widehat{\bm{u}}(\bm{\xi}_m;\bm{c}_{\bm{\alpha}}) - {\bm{u}}(\bm{\xi}_m;\bm{c}_{\bm{\alpha}})\right),
\end{align}
where $\bm{*}^H$ denotes the conjugate transpose of $\bm{*}$.  This yields the solution
\begin{align}
	\widehat{C} = \left(\Psi^H\Psi\right)^{-1}\Psi^HU,
\end{align}
where
\begin{align}
	C &\equiv \begin{bmatrix} \bm{c}_{\bm{\alpha}_1} & \bm{c}_{\bm{\alpha}_2} & \hdots & \bm{c}_{\bm{\alpha}_P}\end{bmatrix}^T,\\
	\Psi &\equiv \begin{bmatrix} \psi_{\bm{\alpha}_1}(\bm{\xi}_1) & \hdots & \psi_{\bm{\alpha}_P}(\bm{\xi}_1) \\ \psi_{\bm{\alpha}_1}(\bm{\xi}_2) & \hdots & \psi_{\bm{\alpha}_P}(\bm{\xi}_2) \\ \vdots & \ddots & \vdots \\ \psi_{\bm{\alpha}_1}(\bm{\xi}_M) & \hdots & \psi_{\bm{\alpha}_P}(\bm{\xi}_M)\end{bmatrix},\\
	U &\equiv \begin{bmatrix} \bm{u}(\bm{\xi}_1) & \bm{u}(\bm{\xi}_2) & \hdots & \bm{u}(\bm{\xi}_M) \end{bmatrix}^T.
\end{align}
Note that the matrix $\Psi$ is only a function of the known random input set $\{\bm{\xi}_m\}$, and $U$ is the matrix of $\bm{u}(\bm{\xi}_m)$ values produced via the deterministic ODE solver.  For a least-squares solution, $M>P$.  While not leveraged in this work, methods based on compressive sampling allowing for generating a PCE with $M\ll P$ with the expansion is sparse \cite{doostan_2011}.

\subsection{Separated Representations}

The method of separated representations (SR), also known as canonical decompositions (CANDECOMP) or parallel factor analysis (PARAFAC), is similar to PCEs in that a solution to a stochastic ODE problem is approximated by projecting the solution onto a multi-variate polynomial basis. The formulation of SR, however, is different from a PCE. Instead of a sum along the multi-index, a surrogate based on SR decomposes a multi-variate function into a sum of products of univariate functions
\begin{equation}
\bm{u}(t,\bm{\xi}) \approx \hat{\bm{u}}_{SR}(t,\bm{\xi}) = \sum_{\ell=1}^r s^\ell \prod_{j=1}^d f_j^\ell(\xi_j),
\end{equation}
where $r$ is the rank of the surrogate and $\{f_j^\ell(\xi_j)\}_{\ell=1}^r, \, j = 1, \ldots , d$ are the univariate functions. The weighting coefficients $\{s^\ell\}_{\ell=1}^r$ are such that each $f_j^\ell(\xi_j)$ has unit norm. These univariate functions, or \textit{factors}, are composed via a sum of coefficients and, in the case of this paper, predetermined polynomials of an orthogonal basis. That is
\begin{equation}
f_j^\ell(\xi_j) = \sum_{n=0}^p c_{j,n}^\ell\psi_n(\xi_j)
\end{equation}
where $c_{j,n}^\ell$ are unknown coefficients and $\psi_n(\xi_j), \, n = 0, \ldots, p$ form a polynomial basis as in the PCE formulation, but where $n$ denotes the degree of the univariate polynomial.

SR utilizes least squares regression in order to calculate the unknown coefficients. Unlike PCEs, the methodology of SR solves for one direction at a time, which reduces the overall computation algorithm to a series of linear optimization problems. In this alternating least squares (ALS) approach,
\begin{equation}\label{eq:argmin_c}
\{\bm{c}_k^\ell\}_{\ell=1}^r = \argmin_{\{\bm{c}^\ell_{k}\}_{\ell=1}^{r}}\frac{1}{M}\sum_{m=1}^M\left\langle u(\bm{\xi}_m) - \hat{u}_{SR}(\bm{\xi}_m),u(\bm{\xi}_m) - \hat{u}_{SR}(\bm{\xi}_m)\right\rangle_2,
\end{equation}
where the variable $k$ denotes the direction being solved for and $\bm{c}_k^\ell = \begin{bmatrix} c_{k,0}^\ell & \cdots & c_{k,n}^\ell \end{bmatrix}^T$.  For the solution to (\ref{eq:argmin_c}), we seek to solve the normal equation
\begin{equation}\label{eq:LS_SR}
(A^HA)\,C_k = A^HU.
\end{equation}
In the ALS process, for each direction, we solve for $C_k$ using (\ref{eq:LS_SR}), where the coefficients are organized as
\begin{equation}
C_k = \begin{bmatrix}{\bm{c}_k^1}^{\,T}&\cdots&{\bm{c}_k^r}^{\,T}\end{bmatrix}^T,
\end{equation}
the vector $U$ is the same as when generating a PCE but for only one QOI, and the matrix $A$ is represented in the block format
\begin{equation}\label{eq:blockA}
A = \begin{bmatrix}{{A}_{11}}&\cdots&{{A}_{1r}}\\ \vdots&\ddots&\vdots\\ {{A}_{M1}}&\cdots&{{A}_{Mr}}\end{bmatrix},
\end{equation}
where ${{A}_{m\ell}}\in\mathbb{C}^{p}$ is given by
\begin{equation}\label{eq:blocks}
{{A}_{m\ell}} = s^\ell\begin{bmatrix}\psi_{0}(\xi_{k,m}) &\cdots&\psi_{p}(\xi_{k,m})\end{bmatrix}\prod_{j \neq k} {f^\ell_{j}} \left(\xi_{j,m}\right),
\end{equation}
and $\xi_{j,m}$ is the $j$th component for sample $\bm{\xi}_m$.  For a more thorough explanation of the ALS process, including a formulation for $\bm{u}$ a vector, see \cite{beylkin_2009b}~and~\cite{balducci_2017}. 
%-----------------------------------------------------------------------------------------------------
\subsection{Selection of Basis Functions}\label{sec:basis_functions}

Selection of the basis function influences convergence for a stochastic expansion.  The solution that produces the fastest convergence, as a function of the number of terms in the expansion, leverages basis functions orthogonal with respect to the posterior distribution (e.g., see discussion in \cite{lemaitre_2010} [pp.~35-36]).  In general, this is not practical since the posterior PDF is not known \emph{a priori}. Instead, basis functions orthogonal to the input probability measure are used, i.e.,
\begin{eqnarray}
	\left<\overline{\psi_j(\bm{\xi})},\psi_k(\bm{\xi})\right> = \int \overline{\psi_j(\bm{\xi})}\psi_k(\bm{\xi})\,d\rho(\bm{\xi}) = \Vert\psi_k\Vert^2\,\delta_{jk},\label{eqn:orthogonality}
\end{eqnarray}
where $\overline{*}$ denotes the complex conjugate of $*$, and $\Vert\psi_k\Vert^2=1$ for the case of orthonormal polynomials.  Typically, (\ref{eqn:orthogonality}) is expressed assuming real numbers, but we generalize the property for use later in this paper.  When this property is satisfied for the basis functions in (\ref{eqn:general_pce_equation}), then 
\begin{equation}
\label{eqn:mean_squares_conv}
\mathbb{E}\left[\left(u(t,\cdot)-\widehat{u}(t,\cdot)\right)^2\right]\xrightarrow[]{m.s.} 0, \quad \mathrm{as}\quad p\rightarrow\infty,
\end{equation}
i.e., the approximation $\widehat{u}(t,\xi)$ converges to $u(t,\xi)$ in the mean-squares sense \citep{cameron_1947}.  Basis functions for many common PDFs are well known as part of the Wiener-Askey scheme \citep{xiu_2002a}, e.g., Hermite and Legendre polynomials for Gaussian and uniform distributed random variables, respectively.  To date, polynomials for PDFs common in the area of directional statistics have not yet been identified and applied to uncertainty quantification using stochastic expansions.  When employing such orthogonal polynomials, both PCEs and SR admit analytic solutions for moments of the posterior distribution as a function of basis function inner products (e.g., see~\cite{lemaitre_2010,doostan_2013}).  The next sections discuss directional statistics and the appropriate set of orthogonal polynomials with regards to a given density.

%=============================================================================================
\section{Random Inputs on the Unit Circle}\label{sec:random_inputs_uc}

\subsection{Directional Statistics}

The field of directional statistics defines the fundamental theory to describe circular data.  The subject began with studying distributions on the compass or in time, both of which may be modeled as functions on the unit circle.  Over time, complexity increased to enable the study of probability and statistics in higher dimensions (e.g., see seminal papers \cite{fisher_1953,mardia_1975a,bingham_1974,kent_1982}).  This section presents pertinent properties of distributions on the unit circle and the probability densities of interest in the current work.

The random angle $\lambda$ has a characteristic function
\begin{align}
	\phi_n = \mathbb{E}\left[e^{in\lambda}\right] = \mathbb{E}\left[z^n\right] = \int_{-\pi}^{\pi} z^n \,d\rho(\lambda),\hspace{12pt}n=0,\pm1,\pm2,\ldots \label{eqn:characteristic_fcn}
\end{align}
with the $n$th trigonometric moments ($n\geq0$) 
\begin{align}
	\mathbb{E}\left[\cos(n\lambda)\right] = \mathcal{R}\left(\phi_n\right),\\
	\mathbb{E}\left[\sin(n\lambda)\right] = \mathcal{I}\left(\phi_n\right),
\end{align}
where $\mathcal{R}(z)$ and $\mathcal{I}(z)$ denote the real and imaginary parts of $z$, respectively.  These are analogous to moments of the PDF seen for typical random variables and uniquely characterize any distribution on the circle (e.g., see \cite[p.~26]{mardia_2000}).  For empirically determined quantities from a collection of $N$ samples,
\begin{align}
	\mathbb{E}\left[\cos(n\lambda)\right]  &\approx \dfrac{1}{N}\sum_{m=1}^N\cos(n\lambda_m),\\
	\mathbb{E}\left[\sin(n\lambda)\right] &\approx \dfrac{1}{N}\sum_{m=1}^N\sin(n\lambda_m).
\end{align}

The mean direction and the circular standard deviation provide an analog to the related quantities on the real line and are functions of the first trigonometric moment.  The circular mean of $\lambda$ is 
\begin{align}
	\mu_{\lambda} &= \tan^{-1}\left(\mathcal{I}\left(\phi_1\right) / \mathcal{R}\left(\phi_1\right)\right)\label{eqn:mean_direction}
\end{align}
and the circular standard deviation is
\begin{align}
    \sigma_{\lambda} &= \sqrt{ -2\,\ln\left(|\phi_1|\right)}.\label{eqn:std_direction}
\end{align}
Note, in this paper, there is no notational distinction between the circular and normal mean/standard deviation aside from the variable over which it is defined.  The circular mean and standard deviation may be approximated empirically using $N$ Monte Carlo samples to estimate $\phi_1$.

\subsection{Distributions on the Unit Circle}

There are two analogs of the Gaussian distribution on the unit circle: the wrapped Normal density $\mathcal{WN}(\mu,\sigma^2)$ \cite[pp.~50-51]{mardia_2000} and the von Mises distribution $\mathcal{VM}(\mu,\kappa)$ \citep{vonmises_1918}.  The (one-dimensional) Wrapped Normal Density (WND)
\begin{align}
	\rho_{\mathcal{WN}}(\xi;\mu,\sigma^2) = \dfrac{1}{\sqrt{2\pi\sigma^2}}\sum^\infty_{k=-\infty} \exp\left\{-\frac{1}{2\sigma^2}\left(\xi-\mu+2\pi k\right)^2\right\}\label{eqn:wrapped_gaussian}
\end{align}
is parameterized by the mean direction $\mu$ and the variance-like quantity $\sigma^2$.  
%
%\begin{align}
%\rho = e^{-\sigma^2/2}.
%\end{align}
%
While similar to the familiar normal distribution on the real line, the sum over $k$ ``wraps" the normal distribution around the unit circle and accounts for the angle rollover.  The characteristic function of the wrapped normal distribution is
\begin{align}
	\phi_{n,\mathcal{WN}} = e^{i\mu n}e^{-n^2\sigma^2/2}.
\end{align}
%
%Given the trigonometric moments of the wrapped normal distribution, then (\ref{eqn:wrapped_gaussian}) may instead be expressed as (see \cite[p.~50]{mardia_2000} and \cite[p.~576]{abramawitz_1972})
%%
%\begin{align}
%	\rho_{\mathcal{WN}}(\xi;\mu,\sigma^2) = \dfrac{1}{2\pi} \vartheta_3\left(\dfrac{\xi-\mu}{2},e^{-\sigma^2/2}\right)\label{eqn:wrapped_gaussian_alt}
%\end{align}
%%
%where
%\begin{align}
%	\vartheta_3(\alpha;\beta) = \sum_{n=-\infty}^\infty \beta^{n^2} e^{2in\alpha}.\label{eqn:vartheta_3}
%\end{align}
%%
%Note that the form of $\vartheta_3$ differs slightly from the discussion in \cite{mardia_2000} so that it may be leveraged later in the selection of the basis polynomials.  The definition of $\vartheta_3$ in (\ref{eqn:vartheta_3}) is mathematically equivalent to the form in \cite{abramawitz_1972} (p.~576) and matches the presentation in \cite{atakishiyev_1994b}.

Initially proposed for applications in atomic physics, the von Mises distribution (VMD) is
\begin{align}
	\rho_{\mathcal{VM}}(\xi;\mu,\kappa) = \dfrac{1}{2\pi I_0(\kappa)}\exp\left\{\kappa \cos\left(\xi-\mu\right)\right\},
\end{align}
where $\mu$ is also the mean direction, $\kappa$ describes the concentration of the PDF, and $I_n$ is the modified Bessel function of the first kind and order $n$.  This PDF accounts for angle ambiguity through the cosine in the argument of the exponent.  For the von Mises distribution
\begin{align}
	\phi_{n,\mathcal{VM}} = \dfrac{I_{|n|}(\kappa)}{I_0(\kappa)}e^{in\mu}.
\end{align}

\begin{figure*}[htbp!]
\centering
\includegraphics[trim=4mm 7mm 2mm 0mm, clip, width=0.7\textwidth]{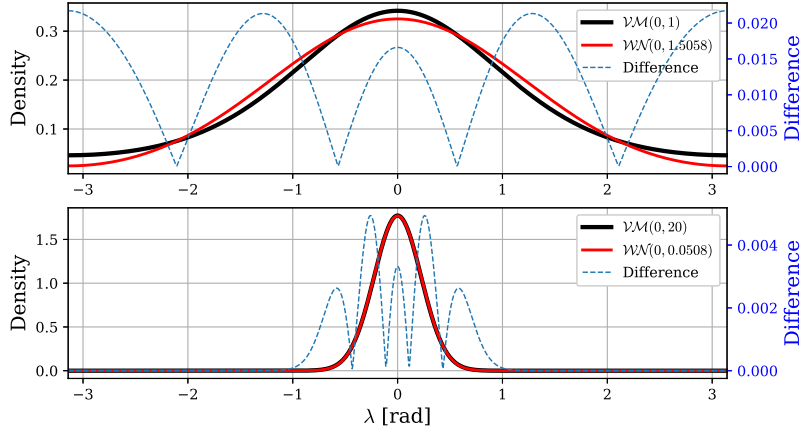}
\caption{Example von Mises and wrapped normal densities}\label{fig:dens_compare}
\end{figure*}

Figure~\ref{fig:dens_compare} presents the WND and VMD with $\mu = 0$ and different values of $\kappa$ and $\sigma^2$.  The indicated value of $\sigma^2$ is based on a nonlinear least squares fit to minimize differences in $\phi_n$, $n=1,\ldots,20$.  In the top case (small $\kappa$ and large $\sigma^2$), the von Mises case has slightly more density in the tails, whereas the wrapped normal is slightly wider in the primary lobe.  The more concentrated case demonstrates the similarity between the two densities for small variance.  The best use of each distribution varies, and \cite{pewsey_2005} discusses the difficulty in discriminating between the two when given samples from one population.  See \cite{collett_1981} for a survey of comparisons between the two densities.  

%Figure~\ref{fig:dens_agreement} illustrates the $\sigma$ produced from a given $\kappa$ using the prescribed least-squares fit.  Differences between the PDFs are quantified using the Kullback-Leibler divergence
%%
%\begin{equation}
%D_{KL}(\rho_{\mathcal{VM}}\,\Vert\rho_{\mathcal{WN}}) = \int_{-\pi}^{\pi} \rho_{\mathcal{VM}}(\xi;\mu,\kappa) \log \dfrac{\rho_{\mathcal{VM}}(\xi;\mu,\kappa)}{\rho_{\mathcal{WN}}(\xi,\mu,\sigma^2)}\,d\xi.
%\end{equation}
%% 
%The KL divergence is computed via $10^5$ uniformly distributed points in $(-\pi,\pi]$.
%
%\begin{figure*}[htbp!]
%\centering
%\includegraphics[trim=0mm 0mm 0mm 0mm, clip, width=0.7\textwidth]{kappa_variance_relationship.eps}
%\caption{WND $\sigma$ value and agreement with VMD as function of $\kappa$.}\label{fig:dens_agreement}
%\end{figure*}

\subsection{Orthogonal Basis Functions}

The following sections describe the polynomials $\psi(z)$ to be used for stochastic expansions with random inputs in $\mathbb{T}$.  The first section describes known polynomials that are orthogonal with respect to the WND and the following section describes generation of polynomials for any PDF with known characteristic moments.

\subsubsection{Orthogonal Polynomials on the Unit Circle}\label{subsubsec:opuc}

Orthogonal polynomials on the Unit Circle (OPUC) serve as the basis functions $\psi_n: \mathbb{T}\rightarrow\mathbb{C}$ for stochastic expansions with random inputs $z$.  While this section provides a brief overview of OPUCs pertinent to this work, we refer the reader to \cite{simon_2004} for a detailed introduction.  The following sections describe specific OPUCs for a given density $\rho(\lambda)$.

The Szeg\H{o} recursion (e.g., see \cite[p.~56]{simon_2004}) enables efficient and stable software implementation of OPUCs.  Let $\psi'_n$ denote an unnormalized form of the polynomial $\psi_n$, and $\psi^{*}_n$ is a (unique) polynomial of degree $n$ that is orthogonal to $z,z^2,\ldots,z^n$ with $\psi^{*}_n(0) = 1$.  Using the Szeg\H{o} recursion,
\begin{eqnarray}
	\psi'_{n+1}(z) &=& z\,\psi'_n(z) - \overline{\eta}_n\psi^{*}_n(z) \\
	\psi^{*}_{n+1}(z) &=& \psi^{*}_n(z) - \eta_n\,z\,\psi'(z)_n\\
	\psi_{n+1}(z) &=& \frac{1}{\Vert \psi'_{n+1}\Vert}\psi'_{n+1}(z)
\end{eqnarray}
where 
\begin{equation}
	\Vert \psi'_{n+1} \Vert^2 = \prod_{i=0}^n (1-|\eta_i|^2),
\end{equation}
and $\eta_i$ are dubbed the Verblunsky coefficients.  These coefficients are unique for a given OPUC defined by measure $\rho(\lambda)$ in Eq.~(\ref{eqn:orthogonality}), and
\begin{equation}
	\eta_n = -\overline{\psi'_{n+1}(0)}.\label{eqn:verblunsky_coeff}
\end{equation}

For the form of the characteristic function in Eq.~(\ref{eqn:characteristic_fcn}), the Toeplitz matrix for a given measure on $\mathbb{T}$ is
\begin{equation}
T_n = \begin{bmatrix} \phi_0 & \phi_{-1} & \hdots & \phi_{-(n-1)} \\ 
				\phi_{1} & \phi_{0} & \hdots & \phi_{-(n-2)} \\  
				\vdots & \vdots & \ddots & \vdots \\ 
				\phi_{n-1} & \phi_{n-2} & \hdots & \phi_{0} \end{bmatrix}.
\end{equation}
As described in \cite[pp.~287-288]{szego_1975}, 
\begin{equation}
	\psi'_n(z) = \left| T_{n} \right|^{-1}\begin{vmatrix} \phi_0 & \phi_{-1} & \hdots & \phi_{-n} \\ 
				\phi_{1} & \phi_{0} & \hdots & \phi_{-(n-1)} \\  
				\vdots & \vdots & \ddots & \vdots \\ 
				\phi_{n-1} & \phi_{n-2} & \hdots & \phi_{-1} \\
				1 & z & \hdots & z^n \end{vmatrix},\label{eqn:numeric_opuc}
\end{equation}
which, when combined with Eq.~(\ref{eqn:verblunsky_coeff}) and computed for $z=0$, allows for computing the Verblunsky coefficients.  While this provides a mathematical framework for generating the OPUC for a given $\rho(\lambda)$ using the associated characteristic function, it can be sensitive to finite-precision arithmetic.  This is especially true for measures highly concentrated on $\mathbb{T}$ (see results and discussion in Section~\ref{sec:numerical_examples}), and are consistent with numeric generation of orthogonal polynomials on the real line (e.g., see \cite{gautschi_1982}).

\subsubsection{Rogers-Szeg\H{o} Polynomials}

\begin{figure*}[htbp!]
\centering
\includegraphics[width=0.8\textwidth]{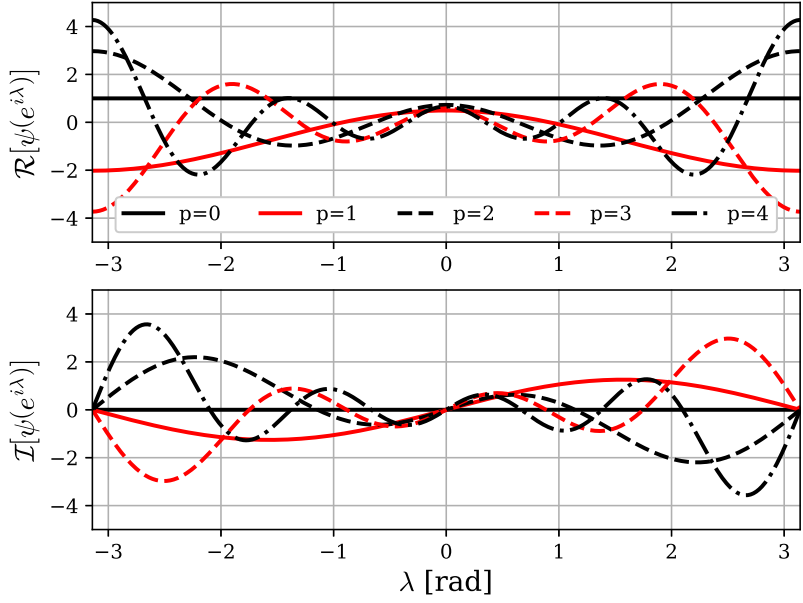}
\caption{Rogers-Szeg\H{o} polynomials of various degree and $q=e^{-1}$}\label{fig:rs-polys}
\end{figure*}

Originally presented by Gabor Szeg\H{o} \citep{szego_1926} and based on the $q$-Hermite polynomials of Leonard Rogers \citep{rogers_1893,rogers_1894}, the Rogers-Szeg\H{o} polynomials are orthogonal with respect to $\rho_{\mathcal{WN}}$.  The (normalized) Rogers-Szeg\H{o} polynomial $\psi_n: \mathbb{T}\rightarrow\mathbb{C}$ of degree $n$ is \citep{atakishiyev_1994b}
%
%\begin{align}
%	\psi_p(z;q) = \sum_{i=0}^{p}(-1)^{i}\begin{bmatrix} p \\ i \end{bmatrix}_q q^{-i/2}\, z^{i}\label{eqn:rogers-szego}
%\end{align}
\begin{align}
	\psi_n(z;q) = (q;q)_n^{-1/2}\,\sum_{j=0}^{n}(-1)^{n+j}\begin{bmatrix} n \\ j \end{bmatrix}_q q^{(n-j)/2}\, z^{j}\label{eqn:rogers-szego}
\end{align}
with parameter $q$ and the $q$-binomial coefficients
\begin{align}
	\begin{bmatrix} n \\ j \end{bmatrix}_q = \dfrac{(q;q)_n}{(q;q)_j(q;q)_{n-j}}, \hspace{18pt}(w;q)_m = \prod^{m-1}_{j=0}(1-w\,q^{j}).
\end{align}
Note that different forms of these polynomials exist in the literature, e.g., see discussion in \cite[p.~87]{simon_2004}.  The Verblunsky coefficients for the Rogers-Szeg\H{o} polynomials are
\begin{equation}
	\eta_n = (-1)^n\,q^{(n+1)/2}.
\end{equation}
These polynomials form an orthogonal basis with regards to the WND measure (e.g., see discussion in \cite{simon_2004,atakishiyev_1994b}) given input argument
\begin{align}
	z = e^{i(\lambda-\mu)},
\end{align}
and parameter
\begin{align}
	q &\equiv e^{-\sigma^2}. \label{eqn:qval}
\end{align}
Note that, via Euler's formula, the Rogers-Szeg\H{o} polynomials are trigonometric polynomials of the random angle $\xi-\mu$.  Figure~\ref{fig:rs-polys} illustrates the real and imaginary components of the Rogers-Szeg\H{o} polynomials for $p=0,\ldots,4$, $\mu=0$, and $q=e^{-1}$.  Note that the polynomials are periodic in $\lambda$.

%To be leveraged in a stochastic expansion with random inputs based on the WND, the Rogers Szeg\H{o} polynomials must be orthogonal with regards to the density measure.  Begin with the (complex) inner product,
%%
%\begin{align}
%	\left<\overline{\psi_n},\psi_m\right> &= \int_{-\pi}^\pi \overline{\psi_n\left(e^{j(\lambda-\mu)};q\right)}\psi_m\left(e^{j(\lambda-\mu)};q\right) \rho_{\mathcal{WN}}\left(\lambda;\mu,\sigma^2\right)\,d\lambda.\label{eqn:rs_normality_angle}
%\end{align}
%%
%Let
%%
%\begin{align}
%	q &\equiv e^{-\sigma^2}, \label{eqn:qval}
%\end{align}
%%
%and recall the relationship between $\vartheta_3$ and $\rho_{\mathcal{WN}}$ in (\ref{eqn:wrapped_gaussian_alt}).  Upon substituting (\ref{eqn:qval}) and (\ref{eqn:wrapped_gaussian_alt}) into (\ref{eqn:rs_normality_angle}), and performing a change in variables to integrate over $z$, then
%%
%\begin{align}
%	\left<\overline{\psi_n},\psi_m\right> &= \dfrac{1}{2\pi j} \oint_{|z|=1} \overline{\psi_n\left(e^{j(\lambda-\mu)};e^{-\sigma^2}\right)}\psi_m(e^{j(\lambda-\mu)};e^{-\sigma^2}) \, \vartheta_3\left(\dfrac{\log z}{2j};e^{-\sigma^2/2}\right) \,\dfrac{dz}{z} \\
%	&= \delta_{nm}
%\end{align}
%%
%where the final result may be found in \cite{atakishiyev_1994b} and adjusted for orthonormal $\psi_n$.  The Rogers-Szeg\H{o} polynomials are orthogonal with regards to the WND and satisfy the condition Eq.~(\ref{eqn:orthogonality}) given random angles $\lambda\sim\mathcal{WN}(\mu,\sigma^2)$.

\section{Stochastic Expansions with Random Inputs on the Unit Circle}\label{sec:stoch_expansion_solution}

For the case of $\xi = \lambda\in\mathbb{T}$, we leverage OPUC, which may be generated using a known $\phi_n$.  Note that, while this work only considers known analytic functions $\phi_n$, approximate polynomials may be generated using approximate values computed via random samples from an unknown population.  The following sections describe the implication of including angle quantities as random inputs or quantities of interest.

\subsection{Random Inputs and QOIs}

This section outlines considerations when employing random inputs and/or QOIs on $\mathbb{T}$.  To aid in understanding future results and clarifying sample generation, methods for simulating data on the unit circle are also included.  Note that, for random angles, this presentation employs $\lambda\in(-\pi,\pi]$.  With appropriate scaling, one may instead use $\lambda\in(-1,1]$ to match other common random inputs.  For ease in presentation, the former is employed in this paper.

Numeric results require simulating data from PDFs on the unit circle.  Simulating samples with distribution $\mathcal{WN}(\mu,\sigma^2)$ requires generating samples from $\mathcal{N}(0,1)$ and mapping them to the unit circle.  Let $\xi'$ be a random sample of the standard Gaussian density $\mathcal{N}(0,1)$.  Then 
\begin{align}
	\lambda(\xi) &= \left(\xi'*\sigma + \mu\right)\mod 2\pi,\label{eqn:simulated_wnd}\\
    \lambda(\xi)  &\sim\mathcal{WN}(\mu,\sigma^2).
\end{align}
This is indicative of the WND being a ``wrapped" Gaussian density around the unit circle.  Random inputs with the WND may be parameterized by $\mathcal{N}(0,1)$ or $\mathcal{WN}(\mu,\sigma^2)$.  Samples of the latter are related to the former via (\ref{eqn:simulated_wnd}).  When describing random samples via $\xi = \xi'$, then Hermite polynomials are the appropriate choice.  For random inputs $\xi = \lambda$, then one should use the Roger-Szeg\H{o} polynomials with input $z = e^{i\xi}$.  Section~\ref{subsec:orbit_cases} presents similarities in the two approaches.  Generation of samples based on the VMD does not allow for a simple transformation of standard random variables to quantities on the unit circle.  Hence, random inputs must be directly generated for $\lambda$, and on the $\xi = \lambda$ approach may be employed.  A simple algorithm to generate samples $\lambda\sim\mathcal{VM}(\mu,\kappa)$ may be found in \cite{best_1979}.

The QOI may be parameterized by $\lambda$ or the circular variable $z = e^{i\lambda}$.  Unless stated otherwise, all angular QOIs in this work are parameterized via $z$.  In such cases, or when using an OPUC for $\psi$, then the coefficients for a given surrogate ($c_{\bm{\alpha}}$ for the PCE and $c^l_{i,n}$ for SR) are in $\mathbb{C}$.  The presentation of methods for generating the surrogate in previous sections is agnostic to real or complex QOIs, and accounting for the differences in software implementation is straightforward.

\subsection{Analytic Moments for PCEs}

Upon generating a surrogate for system response as a function of the random inputs, select characteristics of the posterior density (e.g., mean and variance) are an analytic function of the expansion coefficients.  For the case of QOI $z$, the trigonometric moments, and thus the circular mean and standard deviation, are approximated via coefficients of the expansion.  For the sake of simplicity, this presentation assumes a single QOI (i.e., not a vector $\bm{u}$) and only random inputs on the unit circle.

Trigonometric moments for the case where QOI $u = z$ may be generated using the coefficients of a PCE.  In the case of $\phi_1$,
\begin{align}
	\phi_1 = \mathbb{E}\left[e^{i\lambda}\right] &\approx \mathbb{E}\left[\,\widehat{z}\,\right], \\
	&= \oint_{|\bm{\xi}|=1} \left(\sum_{\bm{\alpha}\in\Lambda_{p,d}} c_{\bm{\alpha}} \psi_{\bm{\alpha}}(\bm{\xi})\right)\,\rho(\bm{\xi}) \,d\bm{\xi},\\
	&= c_{\bm{0}} \\
	&= \mathcal{R}(c_{\bm{0}}) + i\,\mathcal{I}(c_{\bm{0}}),
\end{align}
since $\psi_{\bm{0}}(\bm{\xi}) = 1$ and $\mathbb{E}[\psi_{\bm{\alpha}}(\bm{\xi})] = 0$ for $\bm{\alpha}\neq\bm{0}$.  To solve for the $n=2$ term of the characteristic function,
\begin{align}
	\phi_2 &\approx \mathbb{E}\left[\widehat{z}\,^2\right], \\
	&=  \oint_{|z|=1} \left(\sum_{\bm{\alpha}\in\Lambda_{p,d}} c_{\bm{\alpha}} \psi_{\bm{\alpha}}(\bm{\xi})\right)\left(\sum_{\bm{\alpha}'\in\Lambda_{p,d}} c_{\bm{\alpha}'} \psi_{\bm{\alpha}'}(\bm{\xi})\right)\,\rho(\bm{\xi}) \,d\bm{\xi},\\
	&=  \sum_{\bm{\alpha}\in\Lambda_{p,d}} c_{\bm{\alpha}}^2,
\end{align}
and the final equality results from orthonormal polynomials $\{\psi_{\bm{\alpha}}\}$.  Note that the equation for the second trigonometric polynomial differs from the second moment about the mean for QOIs on the real line, i.e., the sum is computed over all elements.  Higher moment may be generated in a similar fashion, but will require the computation of polynomial triple products, etc.

Given the first trigonometric moment, then the PCE-estimated circular mean and STD are found via
\begin{align}
	\mu_{u} &= \tan^{-1}\left(\mathcal{I}(c_{\bm{0}})/\mathcal{R}(c_{\bm{0}})\right),\label{eqn:surrogate_mean_direction}\\
    \sigma_u &= \sqrt{-2\ln\left(|c_{\bm{0}}|\right)}.\label{eqn:surrogate_std_direction}
\end{align}
While (\ref{eqn:surrogate_mean_direction}) and (\ref{eqn:surrogate_std_direction}) assume a scalar QOI $u$, these may be extended to vector quantities via element-wise operations.

\subsection{Analytic Moments for SR}
As with PCEs, an SR surrogate is able to produce analytical expressions of the moments of the approximated function. To get the first moments, it is straightforward to show that
\begin{align}
	\phi_1 &\approx \mathbb{E}\left[\widehat{z}\right], \\
    &= \oint_{\vert z\vert = 1} \left(\sum^r_{\ell = 1} s^\ell \, \prod^{d}_{j = 1}   f_j^\ell(\xi_j)\right)\rho(\xi_j) d\xi_j,\\
	&= \sum^r_{\ell = 1} s^\ell \, \prod^{d}_{j = 1} c^\ell_{0,j},\label{eqn:sr_first_trig_moment}
\end{align}
and the orthogonality of $\psi_{n}$ leaves the function solely dependent on the zeroth order coefficients and the normalizing constants. Similarly, when considering the second moment,   
\begin{align}
	\phi_2 & \approx \mathbb{E}\left[\widehat{z}\,^2\right], \\
	&=  \oint_{\vert z\vert = 1} \left(\sum^r_{\ell = 1} s^\ell \, \prod^{d}_{j = 1}   f_j^\ell(\xi_j)\right)\left(\sum^r_{\ell = 1} s^\ell \, \prod^{d}_{j = 1}   f_j^\ell(\xi_j)\right)\rho(\xi_j) d\xi_j,\\
	&=  \sum_{\ell = 1}^r \sum^{r}_{\ell' = 1} s^\ell \, s^{\ell'} \,\prod^{d}_{j = 1}  \left(\sum^p_{n = 0} c^\ell_{n,j}\,c^{\ell'}_{n,j} \right).\label{eqn:sr_second_trig_moment}
\end{align}
The trigonometric moments, circular mean, and circular standard deviation may be generated in a manner similar to the PCE given the SR solution for $\phi_1$.

%==========================================================================================
\section{Numerical Examples}\label{sec:numerical_examples}

The following numeric tests demonstrate the efficacy of using stochastic expansions with a random input on the unit circle.  The first case demonstrates the exponential convergence expected when using the appropriate OPUC in a PCE.  The remaining sections then demonstrate practical application of the approach by propagating orbit-state uncertainty with equinoctial elements.  %While results focus on the use of OPUC in PCEs for the sake of simplicity and ease of presentation, tabulated results include performance with SR.

%-----------------------------------------------------------------------------------------------------
\subsection{Stochastic Expansion Convergence}

This section demonstrates the exponential convergence achieved when using OPUC with random inputs defined on $\mathbb{T}$.  Like the convergence rate demonstrations in \cite{xiu_2002a}, these tests consider the stochastic differential equation
\begin{align}
	\dot{u}(t) = -k(\xi)\,u(t),\hspace{24pt}u(0) = u_0
\end{align}
with random decay-rate coefficient $k(\xi)$ and deterministic solution
\begin{align}
	u(t) = u_0 e^{-k(\xi)\,t}.
\end{align}
For this case, $u_0 = 1$, $k(\xi) = z(\xi) = e^{i\xi}$ where $\xi$ is a random variable on the unit circle, and $t=1$ time unit.  The mean and variance of $u(t)$ are
\begin{align}
	\mu_u &= u_0 \int_{-\pi}^{\pi} e^{-k(\xi)\,t}\, \rho(\xi)\,d\xi,\label{eqn:baseline_intrusive_mean}\\
	\sigma^2_u &=u_0 \int_{-\pi}^{\pi} (u-\mu_u)\overline{(u-\mu_u)}\, \rho(\xi)\,d\xi.\label{eqn:baseline_intrusive_variance}
\end{align}
which, for a given density of $\xi$, are computed numerically via quadrature using 1000 uniformly spaced nodes on the unit circle\footnote{See \texttt{CIRCLE\_RULE} software found at https://people.sc.fsu.edu/$\sim$jburkardt/c\_src/circle\_rule/circle\_rule.html [Accessed April 29, 2018]}.  This achieves accuracy in the baseline mean and variance approaching floating point error.  Note that these are not the circular mean or standard deviation since $u$ is not an angle.  

To assess performance of the OPUC basis while eliminating the selection of $M$ from influencing the result, PCE coefficients are numerically propagated forward to $t$, and surrogate-derived statistics are compared to numeric approximations of Eqs.~(\ref{eqn:baseline_intrusive_mean}) and (\ref{eqn:baseline_intrusive_variance}).  Let $c_{k,\bm{\alpha}}$ and $c_{u,\bm{\beta}}$ denote the PCE coefficients for $k$ and $u$, respectively.  In the following, multi-indices $\bm{\beta}$ and $\bm{\gamma}$ follow the same mathematical definition as $\bm{\alpha}$ but use a different symbol to emphasize their use for difference PCEs.  Direct propagation of the coefficients uses a Galerkin projection (see \cite{xiu_2002a} for details) to produce the system of linear differential equations
\begin{align}
	\dot{c}_{u,\bm{\gamma}} &= -\sum_{\bm{\alpha}\in\Lambda_{p,d}}\,\sum_{\bm{\beta}\in\Lambda_{p,d}} e_{\bm{\beta},\bm{\alpha},\bm{\gamma}} \,c_{k,\bm{\alpha}}\,c_{u,\bm{\beta}},\label{eqn:pce_intrusive_eoms}
\end{align}
where
\begin{align}
	e_{\bm{\beta},\bm{\alpha},\bm{\gamma}} = \left<\psi_{\bm{\beta}}\,\psi_{\bm{\alpha}},\overline{\psi}_{\bm{\gamma}}\right>.
\end{align}
The initial conditions are
\begin{align}
	c_{u,\bm{\beta}} = \begin{cases} 1, & \bm{\beta}=\{0\}, \\ 0, & \bm{\beta} = \{j\},\, j > 0\end{cases}
\end{align}
and the PCE for $k(\xi)$ has coefficients
\begin{align}
	c_{k,\bm{\alpha}} = \begin{cases}
	\eta_{1}, & \bm{\alpha}=\{0\}\\
	\sqrt{ 1 - |\eta_1|^2 },& \bm{\alpha} =\{1\}\\
	0, & \bm{\alpha} = \{j\},\,j>1 
	\end{cases}
\end{align}
which is the analytic solution for a PCE such that $k(\xi) = e^{i\xi}$.  The ODE for the PCE coefficients (see Eq.~(\ref{eqn:pce_intrusive_eoms})) may be used in any Runge-Kutta integrator with sufficient order and step size to achieve required accuracy.  All cases use the common 4th order method with a step size of $0.001$.  The triple product values $e_{\bm{\beta},\bm{\alpha},\bm{\gamma}}$ are pre-computed using quadrature integration with the appropriate polynomials $\psi_n$ and stored for use in propagating the PCE coefficients.

\begin{figure*}
	\centering
	\includegraphics[width=0.7\textwidth]{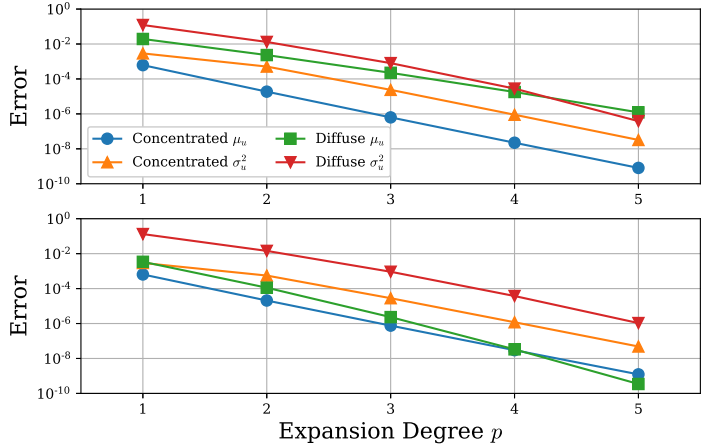}
	\caption{Convergence of the PCE with random inputs from WND (top) and VMD (bottom).}
	\label{fig:convergence_intrusive}
\end{figure*}

Figure~\ref{fig:convergence_intrusive} illustrates the convergence of $\mu_u$ and $\sigma^2_u$ as a function of expansion degree for four different distributions of $k(\xi)$.  The WND and VMD for this case use the parameters identified in Figure~\ref{fig:dens_compare} with ``concentrated" and ``diffuse" referring to larger and smaller values of $\kappa$, respectively.  Error is quantified via
\begin{align}
	\epsilon_{\text{error}} = \left|\dfrac{*_{pce} - *_{ref}}{*_{ref}}\right|,\label{eqn:rel_accuracy}
\end{align}
where $*$ denotes the quantity to characterize, i.e., mean or standard deviation.  Baseline values of the moments for all cases (with enough digits of precision to illustrate differences) are provided in Table~\ref{table:baseline_intrusive_values}.  Even the similar VMD and WND for the concentrated case yield non-trivial differences in moments.  Results imply an exponential convergence in the expansion accuracy as $p$ increases, thereby demonstrating the efficacy of the OPUC for random angles.

\begin{table*}
	\caption{Posterior Moments for intrusive PCE test cases}
	\label{table:baseline_intrusive_values}
	\begin{tabular}{>{\centering\let\newline\\\arraybackslash\hspace{0pt}}m{1in}>{\centering\let\newline\\\arraybackslash\hspace{0pt}}m{0.5in}>{\centering\let\newline\\\arraybackslash\hspace{0pt}}m{1in}>{\centering\let\newline\\\arraybackslash\hspace{0pt}}m{1in}}
	\hline
	Case & Density & $\mu_u$ & $\sigma^2_u$ \\
	\hline
	\multirow{2}{*}{Diffuse} & WND & $0.55345531$ & $0.5423454$ \\
	& VMD & $0.60439010$ & $0.63471260$ \\
	\hline
	\multirow{2}{*}{Concentrated} & WND & $0.36800765$ & $7.2462385\times 10^{-3}$ \\
	& VMD & $0.36801304$ & $7.3271366\times 10^{-3}$
	\end{tabular}
\end{table*}

\begin{figure*}
	\centering
	\includegraphics[trim=0mm 0mm 0mm 0mm, clip, width=0.7\textwidth]{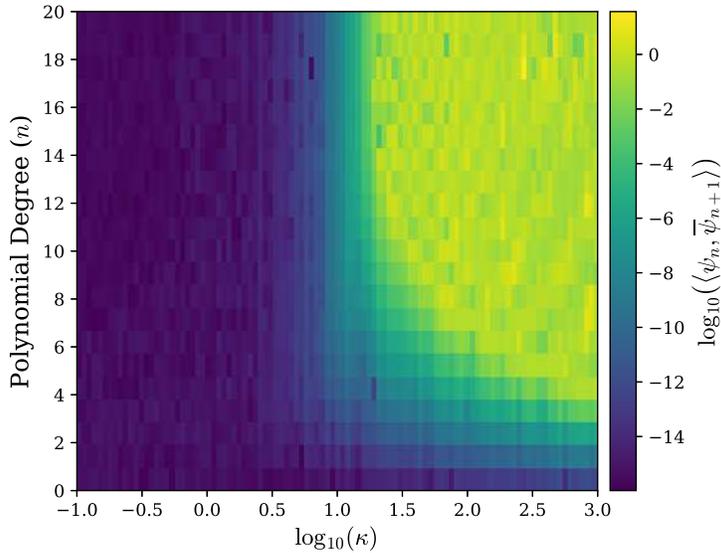}
	\caption{Numeric stability of orthonormal polynomials for von Mises density.}
	\label{fig:vm_poly_stability}
\end{figure*}

Using the numeric procedure in Section~\ref{subsubsec:opuc} to generate the OPUC can suffer from numeric issues for highly concentrated PDFs.  Figure~\ref{fig:vm_poly_stability} demonstrates errors in the OPUC for the VMD quantified as the inner product $\left<\psi_n,\overline{\psi}_{n+1}\right>$ as a function of $\kappa$ and degree $n$.  Each evaluation of the integral (Eq.~(\ref{eqn:orthogonality})) uses $10^{5}$ quadrature points.  These inner products should equal zero, however finite precision arithmetic yields non-zero values for large $\kappa$.  Numeric issues result from the condition number for the matrices in Eq.~(\ref{eqn:numeric_opuc}) when all $\phi_n \approx 1$.  Developing a more numerically stable solution is designated as future work, but previous results for $\kappa = 20$ and $n\leq 5$ still demonstrate solution convergence for that case.  Note that polynomials with analytic solutions for the Verblunsky coefficients, such as the Rogers-Szeg\H{o} polynomials, do not suffer from this issue.  

%-----------------------------------------------------------------------------------------------------
\subsection{Earth Orbit Cases}\label{subsec:orbit_cases}

Of particular interest in this work is propagation of orbit-state uncertainty when using equinoctial elements.  These elements include five quantities ($a$, $h_e$, $k_e$, $p_e$, $q_e$) defined on $\mathbb{R}^5$ with the sixth being the mean longitude $\lambda$.  Hence, propagation of equinoctial elements requires the presented framework with a mixture of QOIs in $\mathbb{R}$ and $\mathbb{T}$.  A series of test cases are presented, some of which focus on specific considerations when mixing polynomials and QOIs in different spaces.

Except when employing two-body only dynamics in Section~\ref{subsubsec:lambda_only}, all samples are propagated in time via special perturbations and TurboProp \cite{hill_2007}.  In such cases, equinoctial elements are converted to Cartesian position and velocity, propagated forward in time, and converted back to equinoctial elements to serve as the QOIs for a given surrogate.  Propagation employs a Dormand-Prince 8(7)~\cite{prince_1981} embedded Runge-Kutta propagator with a relative tolerance of $10^{-12}$.  The gravitational force model uses $\mu_\oplus = 398600.4415$ km$^3$/s$^2$ and $J_2 = 0.00108248$.  For the sake of simplicity, transformation between the Earth-fixed and inertial coordinate systems is assumed to be a simple rotation about an inertially fixed rotation axis.  While these scenarios are restricted to the main problem in Earth-centered orbit propagation for the sake of simplicity and duplication, previous demonstrations of surrogate methods for uncertainty propagation consider higher-fidelity dynamics \cite{jones_2013a,jones_2015b,balducci_2017}.

\begin{table*}
\centering
\caption{\emph{A priori} PDF parameters for orbit test cases at $t_0$}\label{table:orbit_ics}
\begin{tabular}{p{0.75in}p{0.7in}p{0.7in}p{0.7in}p{0.7in}p{0.7in}p{0.75in}}
Moment & $a$ & $h_e$ & $k_e$ & $p_e$ & $q_e$ & $\lambda$ \\
\hline
\hline
Mean/Loc. & 7444.0 km & -7.071$\times$10$^{-2}$ &  7.071$\times$10$^{-2}$ & 7.071$\times$10$^{-1}$ & 7.071$\times$10$^{-1}$ & 33.59 deg \\
Variance & 20$^{\,2}$ & 10$^{-6}$ & 10$^{-6}$ & 10$^{-6}$ & 10$^{-6}$  & varies \\
\hline
\end{tabular}
\end{table*}

Table~\ref{table:orbit_ics} summarizes the \emph{a priori} mean and standard deviations employed in the following tests.  These are loosely based on a test case in \cite{horwood_2011}, but with eccentricity and inclination increased to $0.1$ and $\sim90$ deg, respectively, to promote increased spatial variations in the gravitational field (due to $J_2$) over a single orbit.  Uncertainty in the equinoctial elements remains unchanged when compared to \cite{horwood_2011}, except where required for a given test.  Initial PDFs for the first six elements are Gaussian with the provided standard deviation.  With propagation times on the order of a day or more, this case yields a non-Gaussian posterior PDF for a initial $\sigma_\lambda = 10^{-2}$ deg.  Hermite polynomials are used as the basis functions for all random inputs corresponding to the five elements defined on $\mathbb{R}$.  The $\sigma^2_\lambda$ and any other deviations from Table~\ref{table:orbit_ics} are described in the appropriate section.  Note that any reference to a Root-Mean-Square (RMS) error in the following tests is based on a comparison of realizations of the QOI via evaluation of the propagator and the PCE when given the same random inputs.

\subsubsection{Orbit State Propagation: $\lambda$-Only Case}\label{subsubsec:lambda_only}

\begin{figure*}
	\centering
	\includegraphics[width=0.6\textwidth]{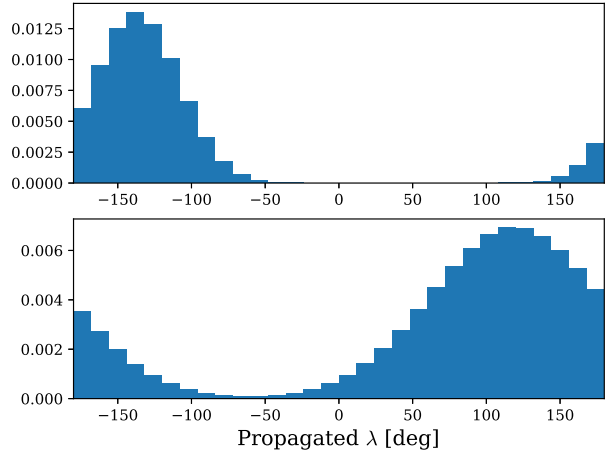}
	\caption{Normalized histogram of propagated $\lambda$ samples after 35 (top) and 70 (bottom) hours.}
	\label{fig:2d_prop_histogram}
\end{figure*}

To focus on propagation of uncertainty for $\lambda$, this section employs two-body only dynamics to propagate the angle:
\begin{align}
	\lambda(t) = \lambda(t_0) + \sqrt{\dfrac{\mu_{\oplus}}{a^3}}(t-t_0),
\end{align}
with random inputs $\xi_1$ and $\xi_2$ corresponding to $a$ and $\lambda$, respectively.  The initial state PDF is consistent with Table~\ref{table:orbit_ics} with $\sigma_\lambda = 10^{-2}$ deg.   Figure~\ref{fig:2d_prop_histogram} depicts the propagated PDF for $\lambda$ after 35.0 and 70.0 hours.  Surrogates presented in this section use $d=2$, $p=10$ and $M=250$.  For each surrogate method, three types of expansions are considered.  Hermite polynomials and random inputs defined by $\mathcal{N}(0,1)$ are considered with QOIs $\lambda$ and $z$.  The third combination considers Roger-Szeg\H{o} polynomials with the circular variable $z$.  A comparable VMD would require $\kappa \approx 3.28\times 10^{7}$.  For reasons of numeric stability, OPUC for the VMD are not considered in this test case.

%\begin{table*}
%\caption{PDF Parameters for Test Cases at $t_0$}\label{table:ics}
%\begin{tabular}{cp{0.75in}p{0.75in}p{0.9in}p{0.9in}p{0.5in}p{0.5in}p{0.75in}}
%\hline\noalign{\smallskip}
%Scenario & Parameter & $a$ & $h$ & $k$ & $p$ & $q$ & $\ell$ \\
%\noalign{\smallskip}\hline\noalign{\smallskip}
%\multirow{2}{*}{LEO} & Mean/Loc. & 7172.571 km & -5.3734$\times$10$^{-4}$ & 9.8269$\times$10$^{-4}$ & -1.1165 & -0.3258 & 30.426 deg \\
%& Variance & 20$^2$ km$^2$ & 10$^{-6}$ & 10$^{-6}$ & 10$^{-6}$ & 10$^{-6}$  & 10$^{-2}$ deg$^2$ \\
%\hline
%\multirow{2}{*}{MOL} & Mean/Loc. & 26,562 km & 0.0 & 0.741 & 0.6176 & 0.0 & 0.0 deg \\
%& Variance & 2$^2$ km$^2$ & 10$^{-6}$ & 10$^{-6}$ & 10$^{-6}$ & 10$^{-6}$  & 10$^{-2}$ deg$^2$ \\
%\noalign{\smallskip}\hline
%\end{tabular}
%\end{table*}

%\begin{table*}
%\caption{LEO Test Case PDF}\label{table:LEO_test_case}
%\begin{tabular}{ccccccc}
%\hline\noalign{\smallskip}
%Moment & $a$ & $h$ & $k$ & $p$ & $q$ & $\lambda$ \\
%\noalign{\smallskip}\hline\noalign{\smallskip}
%Mean ($\overline{*}$) & 6980 km & 0 & 0 & 0 & 0 & 0 rad\\
%Variance ($\sigma_*$) & 400 km$^2$ & 10$^{-6}$ & 10$^{-6}$ & 10$^{-6}$ & 10$^{-6}$ & (10$^{-2}\frac{\pi}{180}$)$^2$ rad \\
%\noalign{\smallskip}\hline
%\end{tabular}
%\end{table*}

\begin{table*}
\caption{Surrogate performance for $\lambda$-only case with $t_f = 35$ hours} \label{table:performance_2d_LEO}
\begin{tabular}{ccccccc}
\hline\noalign{\smallskip}
Method & Basis & QOI & Rel.~Error Mean & Rel.~Error STD & RMS Error [deg] \\
\noalign{\smallskip}\hline\noalign{\smallskip}
\multirow{3}{*}{PCE} & Hermite & $\lambda$ & $2.9\times10^{-2}$ & $5.7\times10^{-1}$ & $45.32$\\
 & Hermite & $z$ & $3.0\times10^{-5}$ & $1.0\times10^{-4}$ & $2.398\times10^{-5}$\\
 & Rogers-Szeg\H{o} & $z$ & $3.0\times10^{-5}$ & $1.0\times10^{-4}$ & $2.399\times10^{-5}$\\
\noalign{\smallskip}\hline\noalign{\smallskip}
\multirow{2}{*}{SR} & Hermite & $z$ & $3.0\times10^{-5}$ & $1.0\times10^{-4}$ & $1.953\times10^{-5}$ \\
& Rogers-Szeg\H{o} & $z$ & $3.0\times10^{-5}$ & $1.0\times10^{-4}$ & $1.953\times10^{-5}$ \\
\noalign{\smallskip}\hline
\end{tabular}
\end{table*}

Table~\ref{table:performance_2d_LEO} presents the performance of the surrogates when approximating the PDF.  The baseline circular mean and standard deviation using 10$^7$ Monte Carlo samples are approximately $-136.89^\circ$ and $28.60^\circ$, respectively.  Relative error was defined previously in Eq.~(\ref{eqn:rel_accuracy}).  For the surrogates leveraging the QOI $z$, mean direction and the standard deviation may be generated using Eq.~(\ref{eqn:surrogate_mean_direction}) and Eq.~(\ref{eqn:surrogate_std_direction}).  Since no analytic solution (as a function of the expansion coefficients) is available for the mean direction and standard deviation given QOI $\lambda$, we use random samples of the surrogate (with the same random inputs used to generate the baseline solution) to compute similar values.  Results demonstrate at least a $\times$1000 improvement in the approximate mean and standard deviation when using $z$, and greater improvement in the RMS error.  The results for different basis functions (Hermite and Rogers-Szeg\H{o}) with QOI $z$ exhibit only slight differences in the RMS error for the PCE surrogates.  Otherwise, no significant difference is exhibited.

\begin{figure*}
	\centering
	\includegraphics[width=0.6\textwidth]{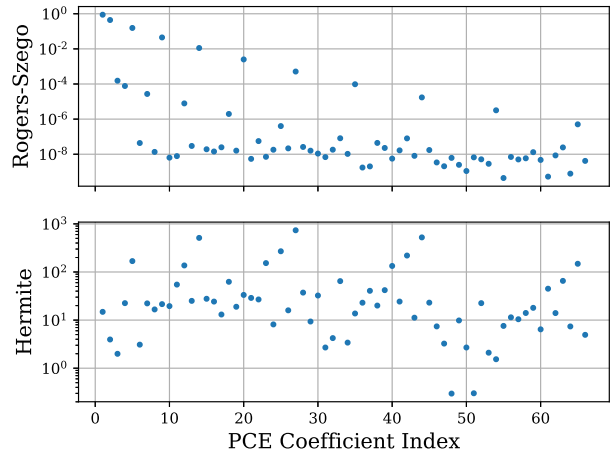}
	\caption{PCE coefficients $|c_{\bm{\alpha}}|$ for the $\lambda$-only case with Rogers-Szeg\H{o} and Hermite polynomials.}
	\label{fig:2dleo_pce_coefficients}
\end{figure*}

Figure~\ref{fig:2dleo_pce_coefficients} illustrates the effect that QOI selection has on the coefficients of the PCE.  When using Rogers-Szeg\H{o} polynomials with QOI $z$, the PCE is converging as the number of terms increases.  However, the Hermite polynomial with $\lambda$ case is failing to converge, there by yielding the poor results seed in Table~\ref{table:performance_2d_LEO}.  Note that the Hermite with QOI $z$ is not included since it duplicates the Rogers-Szeg\H{o} results.  Using the Roger-Szeg\H{o} polynomials demonstrates a clear dependence on a select few PCE terms.  These correspond to terms with high degree for the random input corresponding to $a$, and sensitivity to this input is expected \citep{balducci_2017,horwood_2011}.  The unconverged PCE using the Hermite polynomial fails to provide a similar assessment of solution sensitivity.

\begin{figure*}
	\centering
	\includegraphics[width=0.6\textwidth]{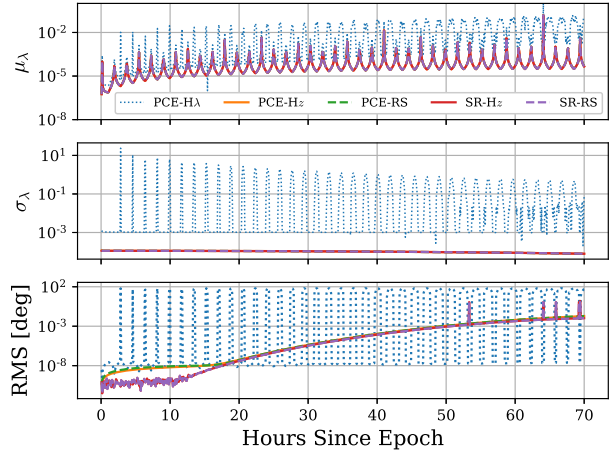}
	\caption{Surrogate accuracy over time for the $\lambda$-only case.}
	\label{fig:2d_orbit_over_time}
\end{figure*}

Figure~\ref{fig:2d_orbit_over_time} presents the performance of the surrogates as a function of time.  Identifiers H$\lambda$ and H$z$ refer to the basis (Hermite) and the QOI ($\lambda$ or $z$).  The circular mean and standard deviation for the H$\lambda$ surrogate uses 10$^6$ random samples of the PCE.  Note that the difference in accuracy for surrogates with QOI $z$ is not visible to the scale of the figure (except as noted below), which is consistent with Table~\ref{table:performance_2d_LEO}.  Performance when using QOI $\lambda$ reduces over time as the propagated PDF becomes more diffuse.  Brief spikes in the H$\lambda$ case depict those times where the PDF is split by the angle boundary (such as the case illustrated in the top image of Figure~\ref{fig:2d_prop_histogram}), and the duration of this reduction in error increases over time.  At later times and for all surrogates, using a QOI $z$ instead of $\lambda$ yields improved performance when compared to the H$\lambda$ case.  The spikes in relative error for $\mu_\lambda$ when using QOI $z$ correspond to times where the mean direction is approximately zero, thereby making the relative error unstable.  The RMS error of $10^6$ independent realizations of the surrogates further demonstrate the improved robustness and accuracy when using a more principled approach to surrogate design.  While error increases over time, which is expected, it may be mitigated by increasing $M$ and $p$.  While a detailed comparison of PCE and SR is outside of the scope of this work, SR exhibits improved RMS accuracy for times less than approximately 15 hours.

\subsubsection{Orbit State Propagation: SMA-Only Case}

\begin{figure*}
	\centering
	\includegraphics[trim=0mm 0mm 0mm 0mm, clip, width=0.6\textwidth]{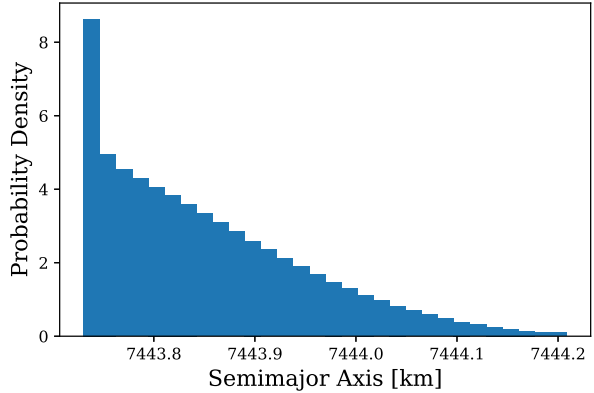}
	\caption{Normalized histogram of propagated samples $a$ given random $\lambda$.}
	\label{fig:taonly_smadist}
\end{figure*}

To demonstrate the effect of random inputs and polynomials in $\mathbb{C}$ on QOIs in $\mathbb{R}$, this section focuses on the effects of uncertainty in $\lambda$ ($d=1$) on $a$.  To allow for $\lambda$ to influence $a$ over time, this case includes the $J_2$ perturbation in orbit propagation.  The sensitivity of $a$ on $\lambda$ will depend on inclination and eccentricity (see \cite[pp.~653-654]{vallado_2007} for discussion), which motivated the change in mean state when compared to test cases in \cite{horwood_2011}.  Note that, for this case, the \emph{a priori} PDF is the WND with $\sigma_\lambda = 5$ deg, and the orbit is propagated for ten orbit periods (approx.~17.75 hours).  All PCE solutions use $M=40$ samples with the Rogers-Szeg\H{o} polynomials, and are compared to a Monte Carlo analysis with $10^{7}$ samples and a resulting $\sigma_a$ precise to approximately four digits.  Figure~\ref{fig:taonly_smadist} presents the posterior (non-Gaussian) marginal PDF for $a$ using these Monte Carlo samples.

\begin{figure*}
	\centering
	\includegraphics[trim=0mm 0mm 0mm 0mm, clip, width=0.6\textwidth]{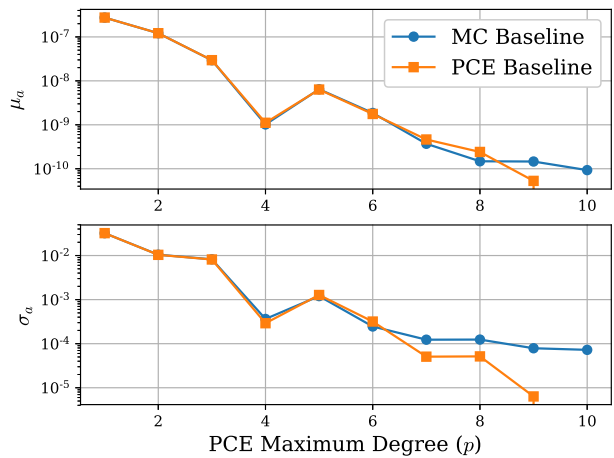}
	\caption{Relative error for PCE-determined mean (top) and standard deviation (bottom).}
	\label{fig:taonly_wn_convergence}
\end{figure*}

Figure~\ref{fig:taonly_wn_convergence} presents the accuracy of the PCE-determined mean and standard deviation for $a$ as a function of $p$.  The Monte Carlo baseline refers to the empirically determined values, while the PCE baseline refers to the $\mu_a$ and $\sigma_a$ from the $p=10$ PCE.  The mean and standard deviation converge to approximately ten and four digits of agreement, respectively, when compared to the independent samples.  This implies agreement between the solutions to the accuracy of the Monte Carlo analysis.  The PCE baseline implies continued convergence to the true value as $p$ increases.  While not presented in the interest of brevity, tests with $M=2000$ samples demonstrate a $10^{-4}$ agreement in $\sigma_a$ for the PCE when $p=3$ and further reductions in differences when compared to the $p=10$ case.

\begin{figure*}
	\centering
	\includegraphics[trim=0mm 0mm 0mm 0mm, clip, width=0.6\textwidth]{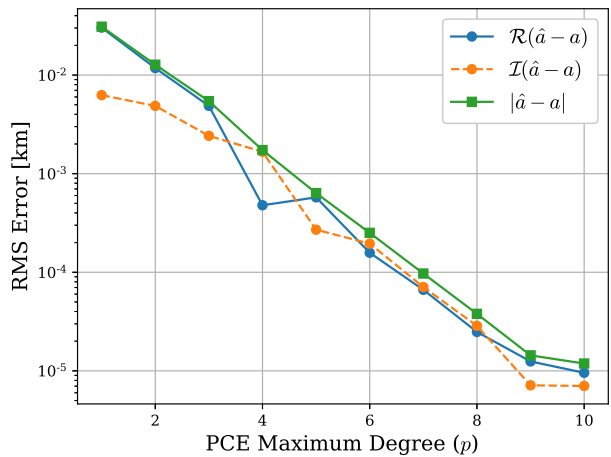}
	\caption{RMS error for PCE-determined $a(\xi)$ as a function of $p$.}
	\label{fig:taonly_wn_smaerror}
\end{figure*}

Figure~\ref{fig:taonly_wn_smaerror} demonstrates the RMS error as a function of $p$.  Since the PCE-produced $\widehat{a}(\bm{\xi})$ is a complex number, this error is quantified three ways as designated in the legend.  For this case, the imaginary and real components of error $\widehat{a} - a$ are comparable in magnitude and decay rapidly with $p$.  While not further explored in this work, using the imaginary component of error as a proxy for solution error is designated for future study.  Normally, estimates of PCE accuracy require cross-validation with independent samples, and a method that does not require additional propagated samples is desired.

%\begin{figure*}
%	\centering
%	\includegraphics[trim=0mm 0mm 0mm 0mm, clip, width=0.6\textwidth]{ta_wn_pcecoeffs.eps}
%	\caption{PCE coefficients for SMA-only case}
%	\label{fig:taonly_wn_coeffs}
%\end{figure*}

\subsubsection{Full Orbit State Propagation}

Propagation of uncertainty for the full equinoctial state is demonstrated in this section for two values of the \emph{a priori} $\sigma_\lambda$.  All elements of the orbit state are considered stochastic, thus $d=6$.  The first case continues to use the relatively small uncertainty $\sigma_\lambda = 10^{-2}$~deg with an \emph{a priori} WND.  To examine performance when the prior PDF for $\lambda$ is a VMD, the second case employs a more diffuse density with $\kappa=30$.  This avoids the numeric issues for large $\kappa$ discussed with Figure~\ref{fig:vm_poly_stability}.  Using the same regression procedure used to identify similar densities in Figure~\ref{fig:dens_compare}, this scenario also employs an \emph{a priori} WND with  $\sigma_\lambda\approx 10.525$ deg.

\begin{figure*}
	\centering
	\includegraphics[trim=0mm 0mm 0mm 0mm, clip, width=0.6\textwidth]{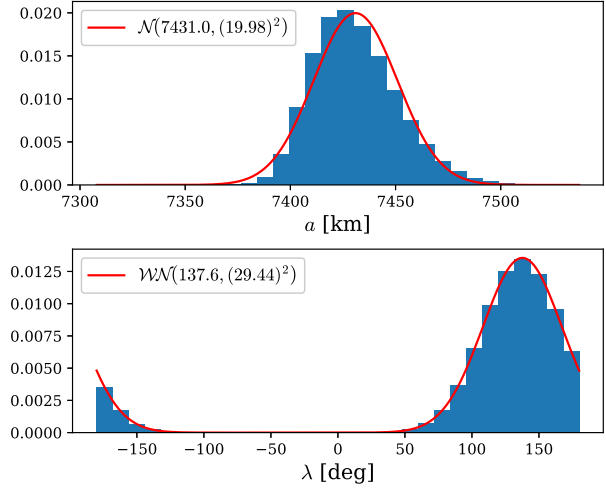}
	\caption{Normalized histogram of propagated $a$ and $\lambda$ components for the $d=6$ case with $\sigma_\lambda = 10^{-2}$ deg.}
	\label{fig:concentratedCase_hist_wn}
\end{figure*}

Figure~\ref{fig:concentratedCase_hist_wn} presents normalized histograms of $a$ and $\lambda$ components of $10^{7}$ propagated samples after $36$ hours with the small initial angle uncertainty.  For the sake of comparison, the figure includes the approximate posterior normal and wrapped normal densities for $a$ and $\lambda$, respectively, based on the empirical $\mu$ and $\sigma$.  The posterior marginal density for $a$ is non-Gaussian, which results from short-period variations induced by the $J_2$ perturbation.  Histograms are not provided for the other four equinoctial elements since the marginal PDFs are approximately Gaussian (to the scale of the resulting figure).

\begin{figure*}
	\centering
	\includegraphics[trim=0mm 0mm 0mm 0mm, clip, width=0.7\textwidth]{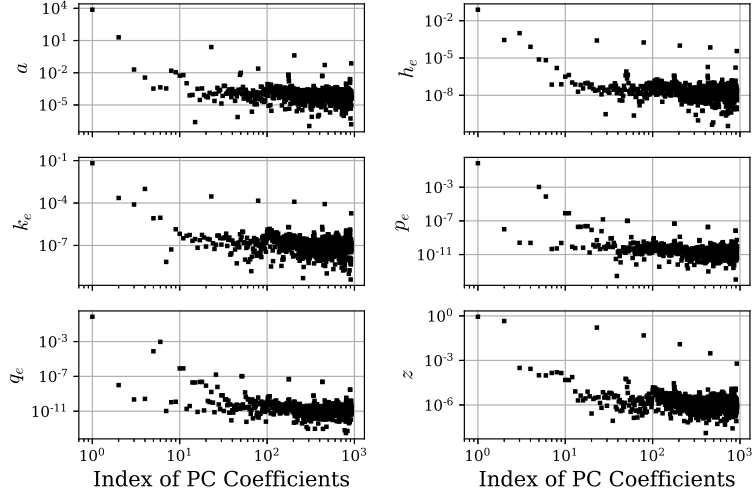}
	\caption{PCE coefficients ($|c_\alpha|$) for the $d=6$ case with $\sigma_\lambda = 10^{-2}$ deg.}
	\label{fig:concentratedCase_pcecoeff_wn}
\end{figure*}

PCE coefficients with the Rogers-Szeg\H{o} polynomials for $\xi_6$ (corresponding to random input $\lambda$) are presented in Figure~\ref{fig:concentratedCase_pcecoeff_wn}.  For this case, $p=6$ and $M=2000$.  A relatively small subset of PCE coefficients influence the propagated PDF, implying that $M$ may be decreased when combined with compressive sampling.  This sparse PCE results from the statistical independence of the equinoctial elements, which was quantified in \cite{balducci_2017} using SR.  In all cases, the PCEs appear to have converged to at least five digits or more.

%\begin{figure*}
%	\centering
%	\includegraphics[trim=0mm 10mm 0mm 20mm, clip, width=0.55\textwidth]{concentratedCase_sensitivity_wn.eps}
%	\caption{Sobol analysis for $d=6$ concentrated case based on PCE solution.}
%	\label{fig:concentratedCase_sensitivity_wn}
%\end{figure*}

\begin{figure*}
	\centering
	\includegraphics[trim=0mm 0mm 0mm 0mm, clip, width=0.7\textwidth]{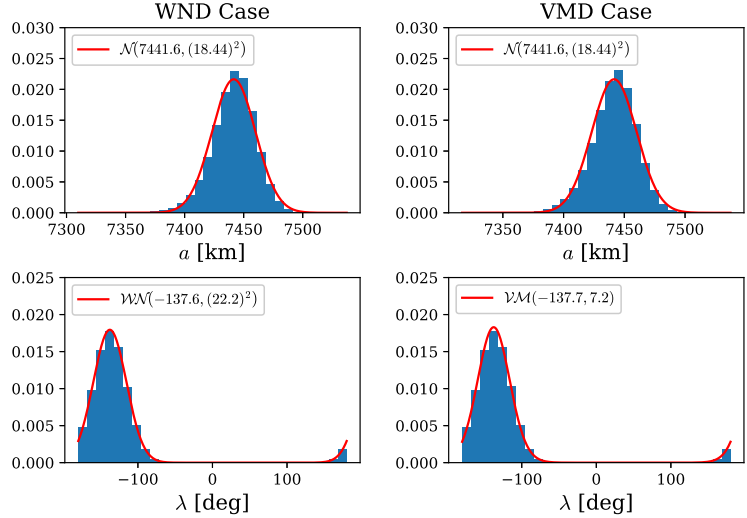}
	\caption{Normalized histogram of propagated $a$ and $\lambda$ components for the $d=6$ diffuse case.}
	\label{fig:diffuseCase_hist}
\end{figure*}

\begin{figure*}
	\centering
	\includegraphics[trim=0mm 0mm 0mm 0mm, clip, width=0.7\textwidth]{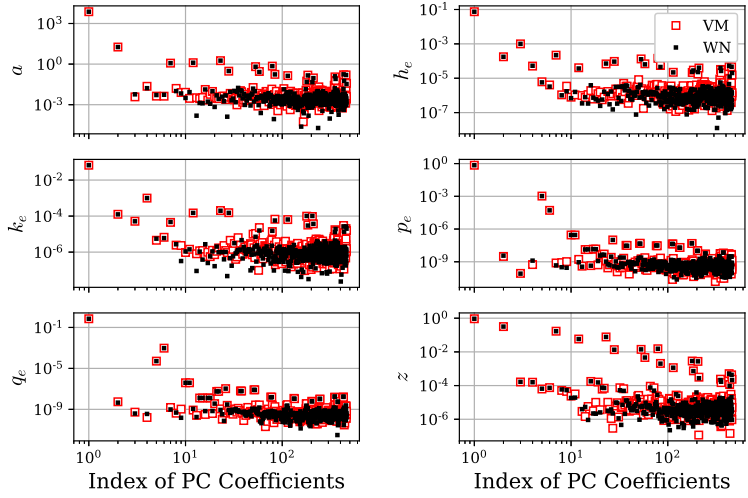}
	\caption{PCE coefficients ($|c_\alpha|$) for the $d=6$ diffuse case.}
	\label{fig:diffuseCase_pcecoeff}
\end{figure*}

Figures~\ref{fig:diffuseCase_hist} and \ref{fig:diffuseCase_pcecoeff} illustrate the posteriori marginal PDFs and PCE coefficients for the case with larger initial circular standard deviation.  Given the more diffuse PDF and its sensitivity to the nonlinear dynamics, results for this case are presented after 24 hours.  As seen in Figure~\ref{fig:diffuseCase_hist}, the propagated $a$ marginal PDF (approximated via $10^7$ samples) is non-Gaussian.  PCEs are generated with $p=5$ and $M=2000$ samples.  Like the more concentrated case, the PCE coefficients imply a sparse approximation may be leveraged.  Slight variations in the posterior solution are seen when comparing the PCE coefficients for the different prior densities for $\lambda$.  While the degree of convergence varies with the QOI, all PCEs appear to be converged to 4 digits or more.

\begin{table*}
\caption{Relative error in $\sigma_u$ for $d=6$ cases when compared to Monte Carlo}\label{table:performance_6d}
\begin{tabular}{>{\centering\let\newline\\\arraybackslash\hspace{0pt}}m{0.5in}|>{\centering\let\newline\\\arraybackslash\hspace{0pt}}m{1in}>{\centering\let\newline\\\arraybackslash\hspace{0pt}}m{1in}>{\centering\let\newline\\\arraybackslash\hspace{0pt}}m{1in}}
\hline
\multirow{2}{*}{QOI ($u$)}& WND & WND & VMD \\
& $\sigma_\lambda = 10^{-2}$ deg & $\sigma_\lambda \approx 10.525$ deg & $\kappa = 30$ \\
%& PCE & SR & PCE & SR & PCE & SR \\
\hline
$a$ & $1.4\times 10^{-4}$ & $8.5\times 10^{-5}$ & $1.9\times 10^{-5}$ \\
$h_e$ & $-1.2\times 10^{-4}$ & $1.7\times 10^{-3}$ & $3.0\times 10^{-3}$ \\
$k_e$ & $-1.8\times 10^{-4}$ & $1.8\times 10^{-3}$ & $9.8\times 10^{-4}$ \\
$p_e$ & $-3.3\times 10^{-4}$ & $-3.3\times 10^{-4}$ & $1.8\times 10^{-5}$ \\
$q_e$ & $-1.3\times 10^{-4}$ & $-1.3\times 10^{-4}$ & $7.1\times 10^{-5}$ \\
$\lambda$ & $1.3\times 10^{-4}$ & $9.7\times 10^{-5}$ & $2.0\times 10^{-5}$
\end{tabular}
\end{table*}

Table~\ref{table:performance_6d} quantifies the accuracy of the surrogate solutions for all cases.  Assessment of the Monte Carlo solution indicates a convergence to approximately four digits of precision for $\sigma_u$ with $10^7$ samples.  As expected, the case with the smaller initial $\sigma_\lambda$ produces better agreement when compared to the Monte Carlo solution.  Consistent with previous results, increases in $p$ and $M$ will demonstrate improved accuracy of the stochastic expansions for propagating uncertainty with equinoctial elements.

%=============================================================================
\section{Conclusions}

Special considerations are required when using a surrogate method such as PCE or SR for uncertainty propagation with circular random inputs or quantities of interest.  Using the circular variable to parameterize directional quantities of interest mitigates the latter issue, but polynomial-based methods require a basis orthogonal with respect to the input probability density functions.  Random inputs described by the wrapped normal density employ the Rogers-Szeg\H{o} polynomials, and a method for numerically generating polynomials allows for generalizing the approach to other densities that are sufficiently diffuse.  Using these polynomials allows for computing directional statistics characterizing the posterior distribution as a function of the coefficients of the stochastic expansion.  Expansions using these polynomials exhibit rapid convergence as a function of degree, and enable propagation of uncertainty in systems with angles as stochastic variables.  This includes propagation of orbit-state uncertainty when using the equinoctial orbital elements.

%\section*{Appendix: Simulating Samples from the VMD}
%
%As presented in \cite{best_1979}  to generate a random samples $\lambda\sim\mathcal{VM}(\mu,\kappa)$, let
%%
%\begin{eqnarray}
%	\tau &=& 1 + \sqrt{1+4\kappa^2},\\
%	\zeta &=& \dfrac{\tau - \sqrt{2\tau}}{2\kappa},\\
%	\theta &=& \dfrac{1+\zeta^2}{2\zeta}.
%\end{eqnarray}
%%
%Use the following procedure to  

\begin{acknowledgements}
Mr.~Balducci's work was funded by the NASA Science and Technology Research Fellowship program, contract NNX15AP41H.  The authors also thank John Kent of the University of Leeds for his discussions on this work.  %The OMP software used in this work was adapted from the solver in SparseLab version 2.0\footnote{Available at https://sparselab.stanford.edu, last accessed August 4, 2017.}
\end{acknowledgements}

% Non-BibTeX users please use
%\begin{thebibliography}{}
%%
%% and use \bibitem to create references. Consult the Instructions
%% for authors for reference list style.
%%
%\bibitem{RefJ}
%% Format for Journal Reference
%Author, Article title, Journal, Volume, page numbers (year)
%% Format for books
%\bibitem{RefB}
%Author, Book title, page numbers. Publisher, place (year)
%% etc
%\end{thebibliography}

\section*{References}
%\bibliographystyle{aiaa_doi}   % Number the references.
%\bibliography{references}   % Use references.bib to resolve the labels.

\begin{thebibliography}{10}
\newcommand{\enquote}[1]{``#1''}
\expandafter\ifx\csname urlstyle\endcsname\relax
  \providecommand{\doi}[1]{doi:\discretionary{}{}{}#1}\else
  \providecommand{\doi}{doi:\discretionary{}{}{}\begingroup
  \urlstyle{rm}\Url}\fi

\bibitem{junkins_1996}
Junkins, J.~L., Akella, M.~R., and Alfriend, K.~T., \enquote{Non-Gaussian Error
  Propagation in Orbital Mechanics,} \emph{Journal of the Astronautical
  Sciences}, Vol.~44, No.~4, 1996, pp.~541--563.

\bibitem{fujimoto_2012b}
Fujimoto, K., Scheeres, D.~J., and Alfriend, K.~T., \enquote{Analytical
  Nonlinear Propagation of Uncertainty in the Two-Body Problem,} \emph{Journal
  of Guidance, Control, and Dynamics}, Vol.~35, No.~2, 2012, pp.~497--509.
  \newline\doi{10.2514/1.54385}.

\bibitem{demars_2013a}
DeMars, K.~J., Bishop, R.~H., and Jah, M.~K., \enquote{Entropy-Based Approach
  for Uncertainty Propagation of Nonlinear Dynamical Systems,} \emph{Journal of
  Guidance, Control, and Dynamics}, Vol.~36, No.~4, 2013, pp. 1047--1057.
  \newline\doi{10.2514/1.58987}.

\bibitem{horwood_2011}
Horwood, J.~T., Aragon, N.~D., and Poore, A.~B., \enquote{Gaussian Sum Filters
  for Space Surveillance: Theory and Simulations,} \emph{Journal of Guidance,
  Control, and Dynamics}, Vol.~34, No.~6, 2011, pp. 1839--1851.

\bibitem{jones_2013a}
Jones, B.~A., Doostan, A., and Born, G.~H., \enquote{Nonlinear Propagation of
  Orbit Uncertainty Using Non-Intrusive Polynomial Chaos,} \emph{Journal of
  Guidance, Control, and Dynamics}, Vol.~36, No.~2, 2013, pp.~430--444.
  \newline\doi{10.2514/1.57599}.

\bibitem{balducci_2017}
Balducci, M., Jones, B.~A., and Doostan, A., \enquote{Orbit uncertainty
  propagation and sensitivity analysis with separated representations,}
  \emph{Celestial Mechanics and Dynamical Astronomy}, Vol.~129, No.~1-2, 2017,
  pp.~105--136. \newline\doi{10.1007/s10569-017-9767-7}.

\bibitem{valli_2013}
Valli, M., Armellin, R., {Di L}izia, P., and Lavagna, M.~R., \enquote{Nonlinear
  Mapping of Uncertainties in Celestial Mechanics,} \emph{Journal of Guidance,
  Control, and Dynamics}, Vol.~36, No.~1, 2013, pp.~48--63.
  \newline\doi{10.2514/1.58068}.

\bibitem{jones_2013b}
Jones, B.~A. and Doostan, A., \enquote{Satellite Collision Probability
  Estimation Using Polynomial Chaos Expansions,} \emph{Advances in Space
  Research}, Vol.~52, No.~11, 2013, pp. 1860--1875.
  \newline\doi{10.1016/j.asr.2013.08.027}.

\bibitem{jones_2015b}
Jones, B.~A., Parrish, N., and Doostan, A., \enquote{Postmaneuver Collision
  Probability Estimation Using Sparse Polynomial Chaos Expansions,}
  \emph{Journal of Guidance, Control, and Dynamics}, Vol.~38, No.~8, 2015, pp.
  1425--1437. \newline\doi{10.2514/1.G000595}.

\bibitem{feldhacker_2016d}
Feldhacker, J.~D., Smith, J., Jones, B.~A., and Doostan, A.,
  \enquote{Multi-Element Trajectory Models for Satellite Tour Missions,} in
  \enquote{{AIAA}/{AAS} {A}strodynamics {S}pecialist {C}onference,} AIAA
  2016-5502, Long Beach, California, 2016.

\bibitem{hintz_2008}
Hintz, G.~R., \enquote{Survey of Orbit Element Sets,} \emph{Journal of
  Guidance, Control, and Dynamics}, Vol.~31, No.~3, 2008, pp.~785--790.

\bibitem{horwood_2014}
Horwood, J.~T. and Poore, A.~B., \enquote{{G}auss von {M}ises Distribution for
  Improved Uncertainty Realism in Space Situational Awareness,}
  \emph{{SIAM/ASA} Journal on Uncertainty Quantification}, Vol.~2, No.~1, 2014,
  pp.~276--304. \newline\doi{10.1137/130917296}.

\bibitem{wiener_1938}
Wiener, N., \enquote{The Homogeneous Chaos,} \emph{American Journal of
  Mathematics}, Vol.~60, No.~4, 1938, pp.~897--936.

\bibitem{ghanem_2002}
Ghanem, R.~G. and Spanos, P.~D., \emph{Stochastic Finite Elements: A Spectral
  Approach}, Dover, New York, 2002.

\bibitem{ghanem_1998}
Ghanem, R.~G. and Dham, S., \enquote{Stochastic Finite Element Analysis for
  Multiphase Flow in Heterogeneous Porous Media,} \emph{Transport in Porous
  Media}, Vol.~32, No.~3, 1998, pp.~239--262.

\bibitem{ghanem_1999a}
Ghanem, R.~G. and Red-Horse, J., \enquote{Propagation of probabilistic
  uncertainty in complex physical systems using a stochastic finite element
  approach,} \emph{Physica D: Nonlinear Phenomena}, Vol.~133, No.~1-4, 1999,
  pp.~137--144. \newline\doi{10.1016/S0167-2789(99)00102-5}.

\bibitem{ghanem_1999}
Ghanem, R.~G., \enquote{Ingredients for a General Purpose Stochastic Finite
  Elements Implementation,} \emph{Computer Methods in Applied Mechanics and
  Engineering}, Vol.~168, No.~1-4, 1999, pp.~19--34.
  \newline\doi{10.1016/S0045-7825(98)00106-6}.

\bibitem{xiu_2002a}
Xiu, D. and Karniadakis, G.~E., \enquote{The {W}iener-{A}skey Polynomial Chaos
  for Stochastic Differential Equations,} \emph{SIAM Journal of Scientific
  Computing}, Vol.~24, No.~2, 2002, pp.~619--644.
  \newline\doi{10.1137/S1064827501387826}.

\bibitem{doostan_2007b}
Doostan, A., Iaccarino, G., and Etemadi, N., \enquote{A least-squares
  approximation of high-dimensional uncertain systems,} Tech. Rep. Annual
  Research Brief, Center for Turbulence Research, Stanford University, 2007.

\bibitem{doostan_2009}
Doostan, A. and Iaccarino, G., \enquote{A least-squares approximation of
  partial differential equations with high-dimensional random inputs,}
  \emph{Journal of Computational Physics}, Vol.~228, No.~12, 2009, pp.
  4332--4345. \newline\doi{10.1016/j.jcp.2009.03.006}.

\bibitem{ma_2009}
Ma, X. and Zabaras, N., \enquote{An adaptive hierarchical sparse grid
  collocation algorithm for the solution of stochastic differential equations,}
  \emph{Journal of Computational Physics}, Vol.~228, No.~8, 2009, pp.
  3084--3113. \newline\doi{10.1016/j.jcp.2009.01.006}.

\bibitem{nouy_2010}
Nouy, A., \enquote{Proper Generalized Decompositions and Separated
  Representations for the Numerical Solution of High Dimensional Stochastic
  Problems,} \emph{Archives of Computational Methods in Engineering}, Vol.~17,
  No.~4, 2010, pp.~403--434. \newline\doi{10.1007/s11831-010-9054-1}.

\bibitem{doostan_2011}
Doostan, A. and Owhadi, H., \enquote{A Non-adapted Sparse Approximation of
  {PDEs} with Stochastic Inputs,} \emph{Journal of Computational Physics},
  Vol.~230, No.~8, 2011, pp. 3015--3034.
  \newline\doi{10.1016/j.jcp.2011.01.002}.

\bibitem{yang_2012}
Yang, X., Choi, M., Lin, G., and Karniadakis, G.~E., \enquote{Adaptive {ANOVA}
  decomposition of stochastic incompressible and compressible flows,}
  \emph{Journal of Computational Physics}, Vol.~231, No.~4, 2012, pp.
  1587--1614. \newline\doi{10.1016/j.jcp.2011.10.028}.

\bibitem{doostan_2013}
Doostan, A., Validi, A., and Iaccarino, G., \enquote{Non-intrusive low-rank
  separated approximation of high-dimensional stochastic models,}
  \emph{Computation Methods in Applied Mechanical Engineering}, Vol.~263, 2013,
  pp.~42--55. \newline\doi{10.1016/j.cma.2013.04.003}.

\bibitem{narayan_2012}
Narayan, A. and Xiu, D., \enquote{Stochastic Collocation Methods on
  Unstructured Grids in High Dimensions via Interpolation,} \emph{SIAM Journal
  of Scientific Computing}, Vol.~34, No.~3, 2012, pp. A1729--A1752.
  \newline\doi{10.1137/110854059}.

\bibitem{ng_2014}
Ng, L. W.~T. and Willcox, K.~E., \enquote{Multifidelity approaches for
  optimization under uncertainty,} \emph{International Journal for Numerical
  Methods in Engineering}, Vol.~100, No.~10, 2014, pp.~746--772.
  \newline\doi{10.1002/nme.4761}.

\bibitem{zhu_2014}
Zhu, X., Narayan, A., and Xiu, D., \enquote{Computational Aspects of Stochastic
  Collocation with Multifidelity Models,} \emph{{SIAM/ASA} Journal on
  Uncertainty Quantification}, Vol.~2, No.~1, 2014, pp.~444--463.
  \newline\doi{10.1137/130949154}.

\bibitem{narayan_2014}
Narayan, A., Gittelson, C., and Xiu, D., \enquote{A Stochastic Collocation
  Algorithm with Multifidelity Models,} \emph{SIAM Journal on Scientific
  Computing}, Vol.~36, No.~2, 2014, pp. A495--A521.
  \newline\doi{10.1137/130929461}.

\bibitem{gerritsma_2010}
Gerritsma, M., van~der Steen, J.-B., Vos, P., and Karniadakis, G.,
  \enquote{Time-dependent generalized polynomial chaos,} \emph{Journal of
  Computational Physics}, Vol.~229, No.~22, 2010, pp. 8333 -- 8363.
  \newline\doi{10.1016/j.jcp.2010.07.020}.

\bibitem{wan_2005}
Wan, X. and Karniadakis, G.~E., \enquote{An adaptive multi-element generalized
  polynomial chaos method for stochastic differential equations,} \emph{Journal
  of Computational Physics}, Vol.~209, No.~2, 2005, pp.~617--642.

\bibitem{wan_2006}
Wan, X. and Karniadakis, G.~E., \enquote{Multi-Element Generalized Polynomial
  Chaos for Arbitrary Probability Measures,} \emph{SIAM Journal on Scientific
  Computing}, Vol.~28, No.~3, 2006, pp.~901--928.
  \newline\doi{10.1137/050627630}.

\bibitem{peng_2015}
Peng, J., \emph{Uncertainty Quantification via Sparse Polynomial Chaos
  Expansion}, {Ph.D.} thesis, University of Colorado Boulder, Boulder, CO,
  2015.

\bibitem{hosder_2006}
Hosder, S., Walters, R.~W., and Perez, R., \enquote{A Non-Intrusive Polynomial
  Chaos Method for Uncertainty Propagation in CFD Simulations,} in
  \enquote{44th AIAA Aerospace Sciences Meeting and Exhibit,} AIAA 2006-891,
  Reno, Nevada, 2006.

\bibitem{beylkin_2009b}
Beylkin, G., Garcke, J., and Mohlenkamp, M.~J., \enquote{Multivariate
  Regression and Machine Learning with Sums of Separable Functions,} \emph{SIAM
  Journal of Scientific Computing}, Vol.~31, No.~3, 2009, pp. 1840--1857.
  \newline\doi{10.1137/070710524}.

\bibitem{lemaitre_2010}
{Le M}a\^{i}tre, O.~P. and Knio, O.~M., \emph{Spectral Methods for Uncertainty
  Quantification with Applications to Computational Fluid Dynamics}, Springer,
  2010.

\bibitem{cameron_1947}
Cameron, R.~H. and Martin, W.~T., \enquote{The Orthogonal Development of
  Non-Linear Functionals in Series of Fourier-Hermite Functionals,}
  \emph{Annals of Mathematics}, Vol.~48, No.~2, 1947, pp.~385--392.
  \newline\doi{10.2307/1969178}.

\bibitem{fisher_1953}
Fisher, R., \enquote{Dispersion on a Sphere,} \emph{Proceedings of the Royal
  Society of London A: Mathematical, Physical and Engineering Sciences},
  Vol.~217, No.~1130, 1953, pp.~295--305. \newline\doi{10.1098/rspa.1953.0064}.

\bibitem{mardia_1975a}
Mardia, K.~V., \enquote{Statistics of Directional Data,} \emph{Journal of the
  Royal Statistical Society. Series B (Methodological)}, Vol.~37, No.~3, 1975,
  pp.~349--393.

\bibitem{bingham_1974}
Bingham, C., \enquote{An Antipodally Symmetric Distribution on the Sphere,}
  \emph{The Annals of Statistics}, Vol.~2, No.~6, 1974, pp. 1201--1225.

\bibitem{kent_1982}
Kent, J.~T., \enquote{The {F}isher-{B}ingham Distribution on the Sphere,}
  \emph{Journal of the Royal Statistical Society. Series B (Methodological)},
  Vol.~44, No.~1, 1982, pp.~71--80.

\bibitem{mardia_2000}
Mardia, K.~V. and Jupp, P.~E., \emph{Directional Statistics}, John Wiley and
  Sons, Ltd., Chichester, England, 2000.

\bibitem{vonmises_1918}
von Mises, R., \enquote{\"{U}ber die Ganzzahligkeit der Atomgewichte und
  Verwandte Fragen,} \emph{Physikalische Zeitschrift}, Vol.~19, 1918,
  pp.~490--500.

\bibitem{pewsey_2005}
Pewsey, A. and Jones, M., \enquote{Discrimination between the von mises and
  wrapped normal distributions: Just how big does the sample size have to be?.}
  \emph{Statistics}, Vol.~39, No.~2, 2005, pp.~81 -- 89.
  \newline\doi{10.1080/02331880500031597}.

\bibitem{collett_1981}
Collett, D. and Lewis, T., \enquote{Discriminating Between the Von Mises and
  Wrapped Normal Distributions,} \emph{Australian Journal of Statistics},
  Vol.~23, No.~1, 1981, pp.~73--79.
  \newline\doi{10.1111/j.1467-842X.1981.tb00763.x}.

\bibitem{simon_2004}
Simon, B., \emph{Orthogonal Polynomials on the Unit Circle, Part I: Classical
  Theory}, American Mathematical Society, Providence, RI, 2004.

\bibitem{szego_1975}
Szeg\"o, G., \emph{Orthogonal Polynomials}, American Mathematical Society,
  Providence, RI, 4th ed., 1975.

\bibitem{gautschi_1982}
Gautschi, W., \enquote{On Generating Orthogonal Polynomials,} \emph{SIAM
  Journal on Scientific and Statistical Computing}, Vol.~3, No.~3, 1982,
  pp.~289--317. \newline\doi{10.1137/0903018}.

\bibitem{szego_1926}
Szeg\H{o}, G., \enquote{Ein Beitrag zur Theorie der Thetafunktionen,}
  \emph{Sitzungsberichte der Preussischen Akademie der Wissenschaften,
  physikalisch-mathematische}, pp.~242--252.

\bibitem{rogers_1893}
Rogers, L.~J., \enquote{Second Memoir on the Expansion of certain Infinite
  Products,} \emph{Proceedings of the London Mathematical Society}, Vol.~s1-25,
  No.~1, 1893, pp.~318--343. \newline\doi{10.1112/plms/s1-25.1.318}.

\bibitem{rogers_1894}
Rogers, L.~J., \enquote{Third Memoir on the Expansion of certain Infinite
  Products,} \emph{Proceedings of the London Mathematical Society}, Vol.~s1-26,
  No.~1, 1894, pp.~15--32. \newline\doi{10.1112/plms/s1-26.1.15}.

\bibitem{atakishiyev_1994b}
Atakishiyev, N.~M. and Nagiyev, S.~M., \enquote{On the {R}ogers-{S}zeg{\H{o}}
  polynomials,} \emph{Journal of Physics A: Mathematical and General}, Vol.~27,
  No.~17, 1994, pp. L611--L615. \newline\doi{10.1088/0305-4470/27/17/003}.

\bibitem{best_1979}
Best, D.~J. and Fisher, N.~I., \enquote{Efficient Simulation of the von Mises
  Distribution,} \emph{Journal of the Royal Statistical Society. Series C
  (Applied Statistics)}, Vol.~28, No.~2, 1979, pp.~152--157.
  \newline\doi{10.2307/2346732}.

\bibitem{hill_2007}
Hill, K., \emph{TurboProp Version 3.2}, Colorado Center for Astrodynamics
  Research, University of Colorado at Boulder, 2007.

\bibitem{prince_1981}
Prince, R.~J. and Dormand, J.~R., \enquote{High order embedded Runge-Kutta
  formulae,} \emph{Journal of Computational and Applied Mathematics}, Vol.~7,
  No.~1, 1981, pp.~67--75. \newline\doi{10.1016/0771-050X(81)90010-3}.

\bibitem{vallado_2007}
Vallado, D.~A. and McClain, W.~D., \emph{Fundamentals of Astrodynamics and
  Applications}, Microcosm Press and Springer, Hawthorne, CA and New York, NY,
  3rd ed., 2007.

\end{thebibliography}

\end{document}